\documentclass [11pt]{article} 
\usepackage{amsmath,amsthm,amsfonts,amscd,eucal,latexsym,amssymb} 
\def\sp{\hskip -5pt} 
\def\spa{\hskip -3pt} 
 \def\bF{{\bf F}}  
 
\newsymbol\bt 1202           
\def\bC{{\mathbb C}}           

\def\bN{{\mathbb N}} 
 
\def\bR{{\mathbb R}} 
\def\bS{{\mathbb S}}

\def\bZ{{\mathbb Z}} 
\newsymbol\rest 1316         

\def\gD{{\mathfrak D}} 
 
\def\gF{{\mathfrak F}} 
 
\def\gH{{\mathfrak H}}

\def\beq{\begin{eqnarray}}
\def\eeq{\end{eqnarray}}

\def\ka{\kappa}

\def\si{\sigma}

\begin{document} 
 
\hfill{\sl UTF 451/UTM 640} 
\par 
\bigskip 
\par 
\rm

 
\par 
\bigskip 
\LARGE 
\noindent
{\bf Holography and $SL(2,\bR)$ symmetry in $2D$ Rindler spacetime} 
\bigskip 
\par 
\rm 
\normalsize 
 
 
\large 
\noindent {\bf Valter Moretti$^{1,2,a}$} and {\bf Nicola Pinamonti$^{b}$} \\
Department of Mathematics, Faculty of Science,
University of Trento, \\ via Sommarive 14, 
I-38050 Povo (TN), 
Italy\\\par 
\smallskip\noindent 
$^1$  I.N.d.A.M.,  Istituto Nazionale di Alta Matematica ``F.Severi'',  unit\`a locale  di Trento\\ 
$^2$  I.N.F.N., Istituto Nazionale di Fisica Nucleare,  Gruppo Collegato di Trento\\
$^a$ E-mail: moretti@science.unitn.it,  $^b$ E-mail: pinamont@science.unitn.it\\ 

\rm\large\large

\rm\normalsize 

\rm\normalsize


\par 
\bigskip 
\par 
\hfill{\sl April-July  2003} 
\par 
\medskip 
\par\rm 
 
\noindent 
{\bf Abstract:}
It is shown that it is possible to define quantum field theory of a massless scalar free field on
the Killing horizon of a $2D$-Rindler spacetime. Free quantum field theory on the horizon enjoys diffeomorphism
invariance and turns out to be  unitarily and algebraically equivalent to the analogous theory of a scalar
field propagating inside Rindler spacetime, nomatter the value of the mass of the field in the bulk. More precisely, 
there exists a unitary transformation that realizes the bulk-boundary correspondence under an
appropriate choice for Fock representation spaces. 
Secondly, the found correspondence is a subcase of  an analogous algebraic correspondence described by injective 
$*$-homomorphisms of the abstract algebras
of observables generated by abstract quantum free-field operators.  These field operators are smeared with 
suitable test functions in the bulk and exact 1-forms on the horizon.
In this sense  the correspondence is independent from the chosen vacua. 
It is proven that, under that correspondence the ``hidden'' $SL(2,\bR)$ quantum symmetry found in a previous work gets a clear geometric meaning, it being associated with a group of 
diffeomorphisms of the horizon itself.   
\newpage
 
\section{Introduction.} 
This paper deals with some holographic properties of quantum field theory in a  manifold 
that admits a (Killing) horizon. The holographic correspondence holds between
QFT in the manifold and QFT suitably defined on the horizon itself.
It is shown that these holographic properties enjoy a nice interplay with
the  hidden $SL(2,\bR)$ symmetry
found in \cite{mopi02}.\\
 In the context  of the problem of the microscopic origin of black-hole entropy,
 holographic principle \cite{thoo93, thoo95, suss95} arose by the idea  
that gravity near the horizon should be described by a low dimensional theory with a higher dimensional group
of symmetry. 
On the other hand, in a very famous paper, Brown and Henneaux \cite{brhe86} described the entropy of an
asymptotically $AdS_3$ black hole in terms of diffeomorphisms preserving the space time structure at spatial
infinity. 
After that, the correspondence between quantum field theories of different dimensions was conjectured by 
Maldacena in his celebrated work about $AdS/Cft$ correspondence \cite{mald98}. Using the machinery of 
string theory, he argued that there is a correspondence between quantum field theory in
a, asymptotically $AdS$,  $d+1$ dimensional spacetime  -- the ``bulk''-- and a conformal theory in a $d$
dimensional manifold -- the (conformal) ``boundary'' at spacelike infinity --.  
Afterwards Witten \cite{witt98} described the above correspondence  in terms of relations
of observables of the two theories. See also the work \cite{gkp98} for further details.
The correspondence in the two dimensional case was studied in \cite{mms98}.
Results arisen by those works were proven rigorously by Rehren for free quantum fields in a $AdS$ background,
exstablishing the existence  
of a correspondence between bulk observables and boundary observables (usually called algebraic holography)
without employing 
string technology \cite{rehr00a,rehr00b}. 
Finally, Strominger \cite{stro01} proposed to enlarge the found results by showing that there
is an analogous correspondence between $dS$ space and a possible conformal field theory on its timelike
boundary. In another work \cite{saso01}, making use of the optical metric, the near horizon limit of a massless
theory in Schwarzschild-like spacetime has been intepreted as a theory in a asymptotic $AdS$ spacetime
giving rise to holographic properties. \\ 
A crucial point to explain the holographic correspondence in Rehren's version is that, in  $AdS_{d+1}$ space, the 
conformal group which 
acts in $d$ dimensions can be realized as the group of the isometries of the $AdS_{d+1}$ bulk.
In this way,  from a pure geometric point of view, the nature of the bulk-boundary correspondence has a
straightforward explanation. 
This is not the case of manifolds with bifurcate Killing horizon
as Kruskal and Minkowski spacetimes.
In Schwarzschild spacetime embedded in Kruskal manifold, the proper boundary relevant to
state holographic theorems  (dropping the boundary at
infinity) seems to be  made of the event horizon of the black hole. Obviously a first and intriguing problem  is the definition of a quantum field theory
on a manifold -- as an (event or Killing) horizon -- whose metric is degenerate. This 
problem is considered in this paper among other related issues.
To approach the general issue in the simplest version,  we notice that two-dimensional Rindler spacetime embedded in Minkowski spacetime approximates the nontrivial part
of the spacetime structure  near a bifurcate horizon as that of a Schwartzschild black hole embedded in Kruskal spacetime.
The remaining transverse manifold is not so relevant in several interesting quantum effects as Hawking's radiation and 
it seems that it can be dropped in the simplest approximation.
In that context,  we have argued in a recent work \cite{mopi02} that free quantum field theory  in
two-dimensional Rindler space  presents a ``hidden'' $SL(2,\bR)$ symmetry. In other words the theory turns out
to be invariant under a unitary representation of $SL(2,\bR)$ but such a quantum symmetry cannot be induced
by the geometric background which enjoys a different group of isometries. 
$SL(2,\bR)$ is the group of symmetry of the zero-dimensional conformal field theory in the sense of \cite{DFF},
so,  as for the case of $AdS$ spacetime, it suggests the existence of a possible correspondence between quantum field
theory in 
Rindler space  and a conformal field theory defined on its (Killing) horizon. 
In fact, as it is shown within this work,
the found hidden symmetry becomes manifest when one examines, after
an appropriate definition,  quantum field theory on the (Killing) horizon. 
That theory enjoys diffeomorphism invariance and 
the $SL(2,\bR)$ symmetry represents, in the quantum context, the 
geometric invariance of the theory under a little group of diffeomorphisms of the horizon. (We stress that
invariance under isometries make not sense since the metric is degenerate.)
We address to section 3 for thecnical details concerning the structure of quantum field theory on the horizon
that, in a sense, is the limit of the bulk theory 
toward the horizon. We only say
here that, generalizing the symplectic approach valid in the bulk,  
the theory can be built up by defining a suitable quantum field operator smeared with exact $1$-forms, which are 
defined on the  horizon, to assure the invariance under diffeomorphisms; 
moreover the causal propagator (which involves bosonic commutation rules)
is naturally defined in spite of the absence,
shared with other holographic
approaches in other contexts,
 of any natural evolution equation. 
The appearance of a manifest quantum $SL(2,\bR)$ symmetry on the horizon  is only a part of the results established in this paper. 
In fact, the manifest $SL(2,\bR)$ symmetry on the horizon is a nothing but a simple result 
which follows form a holographic boundary-bulk
correspondence established in this paper for $2D$ Rindler spacetime either in terms of unitary equivalences 
and in terms of $*$-algebra homomorphisms of free field observables.
 This operator
algebra has a clear geometric interpretation in terms of vector fields defined on the horizon
and  generating 
the group of (orientation preserving) diffeomorphisms of the horizon itself.
Some overlap with our results could be present in the literature. Guido, Longo, Roberts and Verch
\cite{GLRV01} discussed from a general point of view $SL(2,\bR)$ covariant local  QFT defined 
on a bifurcate Killing horizon and obtained by restriction on the horizon of the net of local (Von Neumann) observables (referred to a Hadamard state with respect to the Killing field) given
in the manifold.
Along a similar theme, Schroer and Wiesbrock \cite{SW00} have studied the relationship between QFTs on horizons
and QFTs on the ambient spacetime. They even use the term ``hidden
symmetry" a sense similar as we do here and we done in \cite{mopi02}. In related follow-up works by Schroer \cite{S01}
 and by Schroer and Fassarella \cite{SF01} the relation to holography 
and diffeomorphism covariance is also discussed.\\

This work is organized as follows: next section is devoted to review and briefly improve a few results established in \cite{mopi02}, also giving rigorous proofs,
concerning hidden $SL(2,\bR)$ symmetry for a free quantum scalar field propagating in $2D$ Rindler spacetime.
In the third section we present the main achievement of this work: We build up a quantum field theory for a massless scalar 
field on the
horizon which, in a sense, is the limit toward the horizon of the analogous  theory developed for a (also massive) field propagating 
in the bulk. 
Moreover we show that any free quantum field theories in the bulk and on the horizon are unitarily
and algebraically equivalent (nomatter the value of the mass). In other words there exists a unitary transformation
that realizes the bulk-boundary correspondence upon an appropriate choice for Fock representation spaces. 
In particular, the vacuum expectation values of observables of the free-field theory 
are invariant under the unitary equivalence. 
Actually, as we said, the found correspondence is valid in an algebraic sense too, i.e. it is described by injective 
$*$-homomorphisms of the abstract algebras
of observables constructed by products of free-field operators smeared by suitable test functions/$1$-forms.
In this sense  the correspondence is independent from the
chosen vacua. In the forth section we show that, as we expected, hidden $SL(2,\bR)$ invariance found in \cite{mopi02}
becomes manifast on the horizon. In a forthcoming work \cite{MP4}  we show that the found manifest
$SL(2,\bR)$ unitary representation defined for the horizon QFT can be extended into a 
full positive-energy unitary Virasoro algebra representation with non vanishing central charge
which represents the Lie algebra of vector fields on the (compactified) Killing horizon.
In the last section we make some remarks and comments on the extension of our result to more complicated
spacetimes containing a bifurcate Killing horizon.

\section{Hidden $SL(2,\bR)$ symmetry near a bifurcate Killing horizon.}

{\bf 2.1}.{\em Rindler space}. In \cite{mopi02}
  we have proven that quantum mechanics in a 2D-spacetime which approximates 
the spacetime near a bifurcate Killing horizon  enjoys  {\em hidden} $SL(2,\bR)$ invariance.
This has been done by using and technically improving some general results on $SL(2,\bR)$ invariance
in quantum mechanics \cite{DFF}.
Let us review part of the results achieved in \cite{mopi02} from the point of view of 
quantum field theory in curved spacetime (essentially in the formulation presented in \cite{KW91,Wald} but using
$*$-algebras instead of $C^*$-algebras). \\

\noindent {\bf Remark}. We illustrate  the  construction of quantum field theory in the considered background
 in some details because the general framework will be useful later in developing quantum field theory on a
horizon and holography.   \\

\noindent Consider a Schwarzschild-like metric
\begin{eqnarray} 
 -A(r)dt\otimes dt +{A^{-1}(r)} {dr\otimes dr} + r^2 d\Sigma^2 \label{first}\:, 
\end{eqnarray} 
where $\Sigma$ denotes angular coordinates. Let $R>0$ denote the radius of the 
black hole.
As the horizon is bifurcate, $A'(R)/2\neq 0$ and we can use the following approximation
in the limit $r\to R$
\begin{eqnarray} 
 -{\kappa}^2 y^2 dt\otimes dt + dy\otimes dy + R^2d\Sigma^2 \label{second}\:, 
\end{eqnarray} 
where $\kappa = A'(R)/2$ and $r=R+A'(R)y^2/4$. Dropping the angular 
part $R^2d\Sigma$, the metric  becomes that of the spacetime   called  {\em (right) Rindler wedge}
 ${\bf R}$  which is part of Minkowski spacetime: 
\begin{eqnarray} 
g_{\bf R} := -{\kappa}^2 y^2 dt\otimes dt + dy\otimes dy  \label{rindler}\:, 
\end{eqnarray} 
with global coordinates $t\in (-\infty,+\infty)$, $y\in (0,+\infty)$.
That spacetime is static \cite{Fulling} with respect to the timelike Killing vector $\partial_t$
and spacelike surfaces at constant $t$. Later we shall make use of {\em Rindler light coordinates}
$u,v\in \bR$ which cover ${\bf R}$ and satisfy
\beq
 u := t-\frac{\log(\ka y)}{\ka}\:\:,\:\: v := t+\frac{\log(\ka y)}{\ka}\:
 \:\:\:\mbox{where}\:\:\:t\in \bR\:,y\in (0,+\infty)\:. \label{lc}
\eeq

\noindent {\bf 2.2}.{\em One-particle Hilbert space}.
As ${\bf R}$ is globally hyperbolic \cite{Fulling}, in particular $t$-constant surfaces are Cauchy surfaces, 
quantum field theory can be implemented without difficulties \cite{Wald}.
There is no guarantee for the validity of the  approach 
to quantum field theory 
for static spacetimes based on the quadratic form induced by the 
stress energy tensor presented in \cite{Wald} 
since $-g_{\bf R}(\partial_t,\partial_t)$ has no positive lower bound. 
However we build up quantum field theory of a real scalar field $\phi$ with mass $m\geq 0$ propagating {\cal S}
in  ${\bf R}$ by following a more direct  stationary-mode-decomposition approach.
In fact, {\em a posteriori} it is possible to show that our procedure
produces the same quantization as that in \cite{Wald}.
The Klein-Gordon equation reads
\begin{eqnarray} 
-\partial^2_t \phi  + {\kappa}^2 \left(y \partial_yy \partial_y  -  
 y^2 m^2 \right) \phi =0.  \label{KG} 
\end{eqnarray} 
If $m>0$,  ${\cal S}$ denotes the vector space
of {\em real wavefunctions}, i.e., $C^\infty$ real solutions $\psi$  which have Cauchy data with compact support on a Cauchy surface.
If $m=0$, (\ref{KG}) reduces to 
\begin{eqnarray} 
 (\partial_t + \kappa y\partial_y) (-\partial_t + \kappa y\partial_y)\phi\:\:
\left( = (-\partial_t + \kappa y\partial_y) (\partial_t + \kappa y\partial_y)\phi \right)\:\: = 0  \label{KGF} 
\end{eqnarray}
and the space of real wavefunctions we want to consider is defined as
${\cal S}:= {\cal S}_{\scriptsize\mbox{out}} + {\cal S}_{\scriptsize\mbox{in}}$ 
where ${\cal S}_{\scriptsize\mbox{out}}$ and  ${\cal S}_{\scriptsize\mbox{in}}$  respectively are the space of real 
$C^\infty$ solutions of 
$(\partial_t + \kappa y\partial_y)\psi =0$ and $(-\partial_t + \kappa y\partial_y)\psi =0$
with compactly-supported Cauchy data. 
The compactness requirement does not depend on the 
Cauchy surface \cite{Wald}. 
There are solutions of (\ref{KG}) with $m=0$
which do not belong to ${\cal S}$
in spite of  having compactly-supported Cauchy data 
({ \it With the notation used in (\ref{scalarproduct}), it is sufficient to fix 
compact-support Cauchy data $\psi,n^\mu \partial_\mu \psi$ on a $t$-constant Cauchy 
surface $\Lambda$ such that $\int_\Lambda \partial_\mu \psi \:n^\mu d\sigma \neq 0$}
).
Define in  ${\cal S}\times {\cal S}$  the following {\em symplectic form} \cite{Wald}, which does not depend on the used spacelike
 Cauchy surface  $\Lambda$  with induced measure  $d\sigma$ and  unit normal vector $n$ pointing toward the
future
\begin{eqnarray} \Omega({\psi},\psi') := \int_{\Lambda} \left(\psi' \nabla^\mu {\psi}
     - {\psi} \nabla^\mu \psi'\right)  
 n_\mu\: d\sigma\:.\label{scalarproduct}\end{eqnarray} 
 The definition is well-behaved  for a pair of complex-valued wavefunctions too, and  also if one of these has 
no compactly-supported Cauchy data.
Equipped with these tools we can define  the one-particle 
Hilbert space ${\cal H}$ associated with the Killing field $\partial_t$. 
To this end, consider the two classes of, $C^\infty({\bf R};\bC)$, $\partial_t$-stationary solutions of (\ref{KG}) 
where $K_a$ is the usual Bessel-McDonald function:
\begin{align}
&\Phi_{E}(t,y) := \sqrt{\frac{2E \sinh(\pi E/\kappa)}{\pi^2\kappa}} K_{iE/\kappa}(my) \frac{e^{-iEt}}{\sqrt{2E}}\label{modes}
\:\:\:\:&\mbox{with $E\in \bR^+$, $\:\:\:\:\:$    if $m>0$}\:,\\
&\Phi_{E}\:^{\stackrel{\mbox{\scriptsize  (in)}} {\mbox{\scriptsize (out)}}}(t,y) := 
\frac{e^{-iE(t \pm \ka^{-1}\ln(\ka y))}}{\sqrt{4\pi E}}\label{modesm=0}\:\:\:\:\:&\mbox{with $E\in \bR^+$,  $\:\:\:\:\:$  if $m=0$}
\:,
\end{align} 
where  $\bR^+:= [0,+\infty)$. 
Modes $\Phi^{\mbox{\scriptsize (in)}}_{E}$ are associated with particles crossing the future horizon 
at $t\to +\infty$,  modes $\Phi^{\mbox{\scriptsize (out)}}_{E}$ 
are associated with particles crossing  the past horizon 
at $t\to -\infty$. \\
We have a pair of proposition whose proof is straightforward by using properties of $K_{ia}$, Fourier transform and
Lebedev transform \cite{lebedev}.\\

\noindent {\bf Proposition 2.1. (Completeness of modes)}.
{\em If $m>0$ and $\psi\in {\cal S}$, the function on $\bR^+$
\begin{eqnarray}
E \mapsto \tilde\psi_{+}(E) 
:= -i\Omega\left(\overline{\Phi_{E}},\psi\right)
\label{decomposition2}\:. 
\end{eqnarray}
satisfies 
$\tilde\psi_{+}(E) = \sqrt{E}g(E)$, and $\overline{\tilde\psi_{+}(E)} = -\sqrt{E}g(-E)$, 
 for some  $g\in {\mathsf S}(\bR;\bC)$ (space of complex-valuated  Schwartz' functions on the whole $\bR$).
 Moreover,
\begin{eqnarray}
\psi(t,y) = \int_{0}^{+\infty} \: \Phi_{E}(t,y) \tilde\psi_+(E)\: dE +
\int_{0}^{+\infty} \: \overline{\Phi_{E}(t,y)\tilde\psi_+(E)}\: dE \:\:\:\:\:\mbox{for $(t,y)\in \bR\times (0,+\infty)$} \label{decomposition}\:. 
\end{eqnarray}
If $m=0$ and $\psi\in {\cal S}$, the functions on $\bR^+$ with $\alpha = \mbox{in, out}$ 
\begin{eqnarray}
E \mapsto \tilde\psi^{(\alpha)}_{+}(E) 
:= -i\Omega\left(\overline{\Phi^{(\alpha)}_{E}},\psi\right)
\label{decomposition2m=0}\:, 
\end{eqnarray}
satisfy  $\tilde\psi^{(\alpha)}_{+}(E)=\sqrt{E}g^{(\alpha)}(E)$, 
$\overline{\tilde\psi^{(\alpha)}_{+}(E)}=\sqrt{E}g^{(\alpha)}(-E)$,  where $g^{(\alpha)}\in {\mathsf S}(\bR;\bC)$.
 Moreover, for $(t,y)\in \bR\times (0,+\infty)$
\begin{eqnarray}
\psi(t,y) = \int_{0}^{+\infty} \sum_{\alpha} \Phi^{(\alpha)}_{E}(t,y) \tilde\psi^{(\alpha)}_+(E)\: dE +
\int_{0}^{+\infty} \sum_{\alpha} \overline{\Phi^{(\alpha)}_{E}(t,y)\tilde\psi^{(\alpha)}_+(E)}\: dE\:. \label{decompositionm=0}
\end{eqnarray}} 

\noindent {\bf Proposition 2.2. (Associated Hilbert spaces)}. {\em If $\psi \in {\cal S}$, 
define the one-to-one associated {\em
positive-frequency 
wavefunction} for $m>0$ and $m=0$ respectively 
 \begin{eqnarray}
\psi_{+}(t,y) := \int_{0}^{+\infty} \: \Phi_{E}(t,y) \tilde\psi_{+}(E)\: dE  \:\:,\:\:
\psi_{+}(t,y) := \int_{0}^{+\infty} \: \sum_{\alpha}\Phi^{(\alpha)}_{E}(t,y) \tilde\psi^{(\alpha)}_{+}(E)\: dE  
\label{decomposition+m=0}\:.
\end{eqnarray}
With that definition, for $\psi_1,\psi_2\in {\cal S}$ it results  $\Omega({\psi_{1+}},\psi_{2+})=0$ whereas
\begin{eqnarray}
\langle \psi_{1+},\psi_{2+} \rangle :=  -i\Omega(\overline{\psi_{1+}},\psi_{2+}) \label{hermite} \:,
\end{eqnarray}
is well-defined (at least on $t$-constant surfaces).
Notice that, at least for $m=0$, positive/negative frequency wavefunctions 
cannot have Cauchy data with compact support
due to known analiticity properties of Fourier transform.
Moreover, respectively for $m>0$ and $m=0$
\begin{eqnarray}
\langle \psi_{1+},\psi_{2+} \rangle = \int_{0}^{+\infty} \overline{\tilde\psi_{1+}(E)} \tilde\psi_{2+}(E) \: dE \:\:,\:\:
\langle \psi_{1+},\psi_{2+} \rangle =  \int_{0}^{+\infty} \sum_\alpha \overline{\tilde\psi^{(\alpha)}_{1+}(E)} \tilde\psi^{(\alpha)}_{2+}(E) \: dE\:. 
\end{eqnarray} The {\em one-particle Hilbert} space ${\cal H}$ is defined as the Hilbert completion 
of the space of finite complex linear combinations of functions $\psi_+$, $\psi\in {\cal S}$, 
equipped with the extension of the scalar product (\ref{hermite}) to complex linear combinations of arguments. It results
${\cal H} \cong L^2(\bR^+,dE)$ if $m>0$ or, if $m=0$,
${\cal H} \cong {\cal H}_{\mbox{\scriptsize (in)}} \oplus {\cal H}_{\mbox{\scriptsize (out)}}$ with
${\cal H}_{\mbox{\scriptsize ($\alpha$)}} \cong L^2(\bR^+,dE)$, $\alpha = \mbox{in,out}$.}\\

\noindent {\bf 2.3}.{\em Quantum field operators: Symplectic approach}.
As usual, the whole quantum field is represented in the symmetrized Fock space ${\gF}({\cal H})$ -- that is  
$\cong {\gF}({\cal H}_{\mbox{\scriptsize (in)}})\otimes {\gF}({\cal H}_{\mbox{\scriptsize (out)}})$ in the massless case --
and referred to a vacuum state $\Psi_{\bf R}$ -- namely $\Psi^{\mbox{\scriptsize (in)}}_{\bf R}\otimes \Psi^{\mbox{\scriptsize (out)}}_{\bf R}$
in the massless case -- said the {\em Rindler vacuum state}. $\Psi_{\bf R}$ does not  belong  Hilbert space 
of Minkowski particles in the sense that quantum field theory in Rindler space
and Minkowski one are not unitarily equivalent \cite{Wald}. 
The  {\em quantum field $\Omega(\:\cdot,\hat \phi)$} associated with the real scalar field $\phi$ in (\ref{KG})
is the linear map \cite{Wald}
\begin{eqnarray}
{\cal S} \ni \psi \mapsto \Omega(\psi, \hat\phi) &:=& ia(\overline{{\psi}_+}) -ia^{\dagger}(\psi_+) \label{PHI}\:, 
\end{eqnarray}
 where $\psi\in {\cal S}$ and $a(\overline{\psi_+})$ and $a^{\dagger}(\psi_+)$
 respectively denote the annihilation (the conjugation being used 
 just the get a liner map $\psi_+ \mapsto
a(\overline{\psi_+})$) and construction operator associated with the one-particle state $\psi_+$.
The right-hand side of (\ref{PHI}) is an essentially-self-adjoint  operator defined in the dense invariant subspace 
${\gF}_0\subset {\gF}({\cal H})$
spanned by all states containing a finite  arbitrarily large number of particles with states 
given by positive-frequency  wavefunctions. 
(\ref{PHI}) is formally equivalent via  (\ref{decomposition})
to the non-rigorous but popular definition
\begin{eqnarray}\hat\phi(x) \mbox{ ``='' } \int_0^{+\infty} \Phi_E(x)a_E + \overline{\Phi_E(x)}a^\dagger_E\:  dE\:.\label{naive}\end{eqnarray}
The given procedure can be generalized to any Klein-Gordon scalar field propagating in a (not necessarily static) 
globally hyperbolic
spacetime provided a suitable vacuum state is given \cite{Wald}. Let ${\cal D}({\bf R})$
denote
 the space of real compactly-supported smooth functions in ${\bf R}$ and $J(A)$ 
  the union of the {\em causal past} and {\em causal future} of a set $A\subset {\bf R}$.
As ${\bf R}$  is globally hyperbolic there is a uniquely determined {\em causal propagator} 
$E: {\cal D}({\bf R}) \to {\cal S}$ of the  Klein-Gordon operator $K$ of the field $\phi$ \cite{Wald}. $E$ enjoys the following properties.
It is linear and surjective, $Ef \in J(\mbox{supp}f)$, $Ef=0$ only if $f=Kg$ for some $g\in 
 {\cal D}({\bf R})$ and $E$ 
satisfies for all $\psi\in {\cal S}$, $f,h\in {\cal D}({\bf R})$
\begin{equation}\int_{\bf R} \psi f \:d\mu_g = \Omega(Ef,\psi) \:\:\mbox{and}\:\:
\int_{\bf R} h(x)(Ef)(x) \: d\mu_g(x) = \Omega(Ef,Eh)
\label{psif}\:, 
\end{equation}
 $\mu_g$ being the measure induced by the metric in ${\bf R}$.
(\ref{psif}) suggests to define \cite{Wald} a  quantum-field operator 
{\em smeared  with functions $f$ of ${\cal D}({\bf R})$} as the linear map
\begin{eqnarray}
f\mapsto \hat\phi(f) := \Omega(Ef, \hat\phi)\:.\label{fsmeared}
\end{eqnarray}
It is possible to smear the field operator
by means of compactly-supported complex-valued  functions, whose space will be denoted by ${\cal D}({\bf R};\bC)$,
simply by defining $\hat\phi(f+ih):= \hat\phi(h) + i \hat \phi(h)$ when $f,h\in {\cal D}({\bf R})$. 
(\ref{fsmeared})  entails \cite{Wald} 
\begin{eqnarray}
[\hat\phi(f), \hat \phi(h)] = -iE(f,h) := -i\int_{\bf R} h(x)(Ef)(x) \: d\mu_g(x)\:, \label{locality}
\end{eqnarray}
that is  the   rigorous version of the formal identity $[\hat\phi(x), \hat \phi(x')] = -iE(x,x')$. 
As a further result \cite{Wald} $[\hat\phi(f), \hat \phi(g)] =0$ if
the supports of $f$ and $g$ are {\em spatially separated}, that is
$\mbox{supp} f\not \subset J(\mbox{supp} g)$ (which is equivalent to $\mbox{supp} g \not \subset J(\mbox{supp} f)$).\\
All that we said holds for $m\geq 0$. Let us specialize to the massless 
case giving further details.   In Rindler light coordinates  (\ref{lc}) the  decomposition ${\cal S}={\cal S}_{\mbox{\scriptsize in}}+ 
{\cal S}_{\mbox{\scriptsize out}}$ (see 2.2) reads, if $\psi \in {\cal S}$,
 $\psi(u,v)= \psi(v)+\psi'(u)$
 where $\psi\in {\cal S}_{\mbox{\scriptsize in}}$ and $\psi' \in {\cal S}_{\mbox{\scriptsize out}}$
 are  compactly supported. Trivial consequences are that $\psi$ vanishes on the past horizon $v\to -\infty$, and 
 $\psi'$ vanishes on the future horizon $u\to +\infty$ (see 3.1)
 and  $\Omega(\psi,\psi')=0$. In the considered case
 \begin{equation}E=E_{\mbox{\scriptsize in}} +E_{\mbox{\scriptsize out}}\:, \label{E+E}\end{equation}
 where, in terms of bi-distributions interpreted as in (\ref{locality}), \begin{equation}E_{\mbox{\scriptsize in}}((u',v'),(u,v))= \frac{1}{4}\mbox{sign}(v-v') \:\:\mbox{whereas}\:\:
 E_{\mbox{\scriptsize out}}((u',v'),(u,v))= \frac{1}{4}\mbox{sign}(u-u')\:.\label{E+E'}\end{equation} 
 The maps 
 $f\mapsto E_{\mbox{\scriptsize in/out}}f$
 from ${\cal D}({\bf R})$ to, respectively, ${\cal S}_{\mbox{\scriptsize in/out}}$ are surjective and 
 $E_{\mbox{\scriptsize in/out}}f=0$ if and only if, respectively, $f=\partial_ug$ or $f=\partial_vg$
 for some $g\in {\cal D}({\bf R})$.\\
 In the Fock space associated with Rindler vacuum $\Psi_{\bf R}$,  we have the natural decomposition 
   \begin{equation}\hat{\phi}(f)=\hat{\phi}_{\mbox{\scriptsize in}}(f) + \hat{\phi}_{\mbox{\scriptsize out}}(f)
  \:\:\:\:\mbox{with}\:\:\:\: \hat{\phi}_{\mbox{\scriptsize in/out}}(f):=\Omega(E_{\mbox{\scriptsize in/out}}f,\hat\phi) 
  \label{decm=0}\end{equation} and the
  the two kinds of field operators commute, i.e.
$[\hat{\phi}_{\mbox{\scriptsize in}}(f),\hat{\phi}_{\mbox{\scriptsize out}}(g)]=i\Omega(E_{\mbox{\scriptsize in}}f,
E_{\mbox{\scriptsize out}}g)=0$. \\

\noindent{\bf 2.4}.{\em $SL(2,\bR)$ symmetry}.
If $m>0$ and thus  ${\cal H}\cong L^2(\bR^+,dE)$, the {\em one-particle (Rindler) Hamiltonian}
$H$ is the self-adjoint operator 
\begin{equation}(Hf)(E) := Ef(E) \:\:\:\: \mbox{with domain ${\cal D}(H) = \{f\in L^2(\bR^+,dE) \:|\: \int_{0}^{+\infty} E^2 |f(E)|^2 dE <+\infty\}$}
\label{H} \:. \end{equation}
If $m=0$ and thus ${\cal H} \cong L^2(\bR^+,dE)\oplus L^2(\bR^+,dE) \cong \bC^2 \otimes L^2(\bR^+,dE) $, the Hamiltonian $H$ reads
  $I\otimes H'$, $H'$ being  the operator defined in the right-hand side of (\ref{H}) 
referred to $L^2(\bR^+,dE)$ and  the identity operator $I$ being referred to $\bC^2$.\\
In \cite{mopi02} it was argued that the one-particle system enjoys invariance under a unitary representation of $SL(2,\bR)$
as consequence of the form of the spectrum of $H$ which is $\sigma(H)= [0,+\infty)$ with no degeneracy for $m\neq 0$ and double degeneracy
if $m=0$. Let us state and prove rigorously some of the statements of \cite{mopi02} in a form which is relevant for 
the remaining part of this work. First of all one has to fix a real constant $\beta>0$ \cite{mopi02}, with the physical dimensions 
 {\em energy}$^{-1}$, that is necessary for dimensional
reasons in defining the relevant domain of operators as it will be clear from the proof of Theorem 2.1. {\em We assume 
to use the same value of $\beta$ throughout this work.}   
Fix reals $k,m>0$ and define the dense subspace ${\cal D}_k\subset {\cal H}\cong  L^2(\bR^+,dE)$ spanned 
by vectors:
\begin{eqnarray}
Z^{(k)}_n(E) := \sqrt{\frac{\Gamma(n-k+1)}{E \: \Gamma(n+k)}}\:e^{-\beta E} 
\left(2\beta E\right)^k
\mathsf{L}^{(2k-1)}_{n-k}\left(2\beta E\right)\:, \:\:\:\:\mbox{$n=k,k+1,\ldots$,}\label{ZEgen}
\end{eqnarray}
where  $\mathsf{L}^{(\alpha)}_p$ are modified Laguerre's polynomials \cite{grads}. 
Notice that ${\cal D}_k\subset {\cal D}(H)$. Moreover, ${\cal D}_k$ is invariant under the linearly-independent symmetric operators
defined on ${\cal D}_k$:
\begin{equation}
H_0 := H\spa\rest_{{\cal D}_k}\:,\:\:\:\:\:
D := -i\left(\frac{1}{2} + E\frac{d \:}{d E}\right)\:, \:\:\:\:\:
C := -\frac{d \:}{d E} E\frac{d \:}{d E} +
\frac{(k-\frac{1}{2})^2}{E}\:.\label{ge3}  
\end{equation} 
which enjoy the commutation relations of the Lie algebra of $SL(2,\bR)$, $sl(2,\bR)$,
\begin{equation} 
[iH_0,iD] = -iH_0 \:,\:\:\:\:\:
[\:iC,iD] =  i C \:, \:\:\:\:\:
[iH_0,iC] = -2iD \:.\label{three} 
\end{equation}
$iH_{0}, iC, iD$ are 
operatorial realizations of the basis elements of  $sl(2,\bR)$
\begin{eqnarray}
 h = \left[
\begin{array}{cc}
  0 &1\\
  0 & 0 
\end{array}
\right]\:,\label{si1}
\:\: c = \left[
\begin{array}{cc}
 0 & 0\\
 -1 & 0 
\end{array}
\right]\:,\label{si2}
\:\: d = \frac{1}{2}\left[
\begin{array}{cc}
1 & 0 \\
0 & -1
\end{array}
\right]\:.\label{si3}
\end{eqnarray}
As a consequence, one expects that operators in (\ref{ge3}) generate a strongly-continuous unitary
representation of $SL(2,\bR)$. 
A complete treatments of the representations of $SL(2,\bR)$ can be found in \cite{barg47,sally,pukansky}.
Let ${\cal L}$ indicate the space of finite
real linear combinations of operators in (\ref{ge3}), let $\rho :  sl(2,\bR) \to i{\cal L}$ be the unique Lie
algebra isomorphism with $\rho(h)=iH_0$, $\rho(c)=iC$, $\rho(d)=iD$
 and let $\widetilde{SL}(2,\bR)$ denote the universal covering of ${SL}(2,\bR)$. \\
 
\noindent {\bf Theorem 2.1}. {\em The operators of ${\cal L}$ are essentially self-adjoint, 
$\overline{H_0}=H$ in particular, and: 

{\bf (a)} ${\cal H}$ is irreducible under the unique unitary strongly-continuous representation of
$\widetilde{SL}(2,\bR)$, $g\mapsto U_g: {\cal H}\to {\cal H}$
such that $U(\exp{(tx)}) =e^{it\:\overline{\rho(x)}}$ for all $x\in sl(2,\bR)$, $t\in \bR$. If (and only if) $k \in \{1/2,1, 3/2,\ldots \}$,
$U$ is a representation of $SL(2,\bR)$ and is faithful only if $k=1/2$. $U$ does not depend on $\beta>0$.

{\bf (b)} $\{U_g\}_{g\in \widetilde{SL}(2,\bR)}$ is a group of {\em symmetries} of the quantum system, that is, 
for every $t\in \bR$ and $g\in \widetilde{SL}(2,\bR)$, there is $g(t)\in \widetilde{SL}(2,\bR)$
 such that
\begin{eqnarray}
 e^{itH} \:U_g \:A\: U_g^\dagger \:e^{-itH}=  U_{g(t)}\: e^{itH}\: A \:e^{-itH} \:U_{g(t)}^\dagger
 \label{invariancetrue}\:,
 \end{eqnarray}
 for every observable (i.e., self-adjoint operator) $A$. If
 $g=\exp(uh+vc+wd)$, with $u,v,w\in \bR$, 
\begin{eqnarray}
 g(t)=\exp((u+tw+t^2v)h+(w+2tv)d+ vc)\:.\label{g(t)}
 \end{eqnarray}
 
 {\bf (c)} For every $t\in \bR$, consider the linearly independent
elements of ${\cal L}$ 
\begin{equation} 
H_0(t) := H_0\:,\:\:\:\:
D(t) := D + tH\:,  \:\:\:\:\:
C(t) := C + 2t D + t^2 H \label{Ct}\:. 
\end{equation} 
The time-dependent observables generated by those operators
 are {\em constants of motion}, i.e.  
\begin{eqnarray}
e^{itH} \: \overline{uH_0(t)+ vC(t) +w D(t)} \: e^{-itH} \:=\: \overline{uH_0+ vC +w D}\:, \:\:\:\:\mbox{for every $t,u,v,w
\in \bR$\:.}\label{trivial2}
\end{eqnarray}}

\noindent {\em Proof}. 
{\bf (a)} $\{Z^{(k)}_n\}_{n=k,k+1,\ldots}$ (\ref{ZEgen}) is a  Hilbert base of 
eigenvectors of the operator $K= \frac{1}{2}(\beta H_0 + \beta^{-1}C)$ \cite{mopi02}.
 Moreover  $X=\beta^2H_0^2+\beta^{-2}C^2+2D^2$ is essentially selfadjoint in ${\cal D}_k$ 
 because $\{Z^{(k)}_n\}_{n=k,k+1,\ldots}$ are analytic vectors for $X$ since
  $X= 4K^2 +cI$ from (\ref{three}) where $c\in \bR$. 
 Since $X$ is  essentially self-adjoint, general results due to Nelson (Theorem 5.2, Corollary 9.1, Lemma 9.1 and Lemma 5.1 
 in \cite{nelson})
 imply that the operators in $i{\cal L}$, are essentially self-adjoint on ${\cal D}_k$ moreover they imply
 the existence and uniqueness of a unitary representation on ${\cal H}$ of the
 unique simply connected group with Lie algebra given by the space generated by $\beta h,\beta^{-1}c,2d$ (that is
 $\widetilde{SL}(2,\bR)$) such that $\overline{\frac{d}{dt}|_{t=0}U(\exp{(tx)})}= i\overline{\rho(x)}$
 for all $x\in sl(2,\bR)$. 
 The derivative in the left-hand side is evaluated in the strong operator topology sense 
 on a suitable subspace $G$ (G\.arding space \cite{nelson}) and gives a restriction of the Stone generator of the 
 strongly continuous unitary one-parameter subgroup $\bR\ni t\mapsto U(\exp{(tx)}))$. As  $G$
 contains a dense set of analytical vectors for the elements of ${\cal L}$  \cite{nelson}, $-i\frac{d}{dt}|_{t=0}U(\exp{(tx)})$ 
 is essentially self-adjoint and thus its closure coincides with the usual Stone generator and $U(\exp{(tx)}) =e^{it\:\overline{\rho(x)}}$.
 As ${\cal D}_k\subset {\cal D}(H)$, the unique self-adjoint
 extension of $H_0$, $\overline{H_0}$, must coincide with $H$ itself.
 Suppose that $P$ is the orthogonal projector onto an invariant subspace for each $U_g$. $P$ must commute with
  $e^{it \overline{K}}$
 in particular. Using Stone's theorem and the fact
 that the spectrum of $K$ is not degenerate, one has that (in strong operator topology sense) 
 $P = \sum_{i\in M} Z^{(k)}_n \langle Z^{(k)}_n, \:\cdot \: \rangle$
 where $M\subset \bN$. Similarly $P$ must commute with every element of ${\cal L}+ i{\cal L}$, 
 $A_{\pm}:= \mp iD + \frac{1}{2}(\beta H_0 - \beta^{-1}C)$ in particular.
 Using the fact that, for every $m,n \in \bN$ with $m>0$, 
 $Z^{(k)}_{n+1}= c_nA_{+}Z^{(k)}_n$ and 
 $Z^{(k)}_{m-1}= c_mA_{-}Z^{(k)}_m$ 
 for some reals  $c_n,c_m>0$
 \cite{mopi02}, one proves that $M=\bN$, that is $P=I$ and so the representation is irreducible.
 The proof of the fact that the representation of $\widetilde{SL}(2,\bR)$  
 reduces to a representation of $SL(2,\bR)$ iff $k\in  \{1/2,1, 3/2,\ldots \}$ 
 and that the representation is faithful only for $k=1/2$ 
 is based on the representation of the subgroup $\{\exp{t(h+c)}\}_{t\in \bR}\subset \widetilde{SL}(2,\bR)$ 
 which is isomorphic to $SO(2)$ and is responsible for the fact that 
 $\widetilde{SL}(2,\bR)$ is multiply connected.
 The proof has been sketched in section 6.2 and the footnote 4 in \cite{mopi02}. 
 The self-adjoint extensions of the elements in ${\cal L}$ do not depend on the value of $\beta>0$
 -- and thus it happens for the representation $U$ itself since every $g\in \widetilde{SL}(2,\bR)$
 is the finite product of elements of some one-parameter subgroups -- 
  because,
 if $\beta'\neq \beta$, there is a subspace ${\cal D}$ containing, with obvious notation, both ${\cal D}_k^{(\beta)}$ 
 and  ${\cal D}_k^{(\beta')}$  where each element of ${\cal L}$ (which is essentially self adjoint on both 
 ${\cal D}_k^{(\beta)}$ 
 and  ${\cal D}_k^{(\beta')}$) determines the same symmetric extension
 nomater if one starts by ${\cal D}_k^{(\beta)}$ or ${\cal D}_k^{(\beta')}$. That extension is essentially self-adjoint since
 ${\cal D}_k^{(\beta)}$ is a dense set of analytic vectors in ${\cal D}$.
 To prove {\bf (b)} and {\bf (c)} notice that $e^{itH}=U(\exp(th))$.
 So,  if $g\in \widetilde{SL}(2,\bR)$,
 $g(t) := \exp(th) g (\exp(th))^{-1}$ fulfils (\ref{invariancetrue}) by application of the representation $U$.
 Define $h(t):=\rho^{-1}(H_0(t))$, 
 $c(t):=\rho^{-1}(C(t))$, $d(t):=\rho^{-1}(D(t))$. These matrices satisfies the commutation rules
  (\ref{three}) for every $t$.
  Using (\ref{three}) and uniqueness theorems for matrix-valued differential equations one   
  gets, for $s,t,u,v,w\in \bR$,
  $$\exp\{th\}\exp\{s(uh(t)+ vc(t) +wd(t))\}(\exp\{th\})^{-1} =
  exp\{s(uh+ vc +wd)\}\:,$$
  which is  (\ref{g(t)}) if $s=1$. 
 Applying $U$ on both sides one gets
 $e^{itH}e^{is\:\overline{uH_0(t)+ vC(t) +w D(t)}} e^{-itH} =
 {e^{is\:\overline{uH_0+ vC +w D}}}\:, \label{trivial}$
which is equivalent to $e^{is\exp\{itH\}\overline{uH_0(t)+ vC(t) +w D(t)}\exp\{-itH\}} =
 {e^{is\:\overline{uH_0+ vC +w D}}}$.
Stone's theorem entails (\ref{trivial2}) by strongly differentiating both sides in $s$.
  $\Box$\\

\noindent The generalization to the case $m=0$ is trivial: The theorem  holds true separately in each 
space $L^2(\bR^+,dE)$ of ${\cal H}\cong L^2(\bR^+,dE) \oplus L^2(\bR^+,dE)$.\\

\noindent {\bf Remarks}.
{\bf (1)} From now on we assume to work in {\em Heisenberg representation}.
Within this picture, by (\ref{trivial2}), $H$, $\overline{C}$, $\overline{D}$ coincide with the Heisenberg evolution of, respectively,
$\overline{H_0(t)}$, $\overline{C(t)}$, $\overline{D(t)}$ at time $t$.  
Moreover, in this picture, $e^{-i\tau H} \psi_+$  must not seen as the time evolution (at time $\tau$) of the  
state $\psi_+$ (given  at time $0$), but it must considered as a different state at all. 
This turns out to be in accordance with the relationship between states and wavefunctions (see 2.5):  $\psi$ and 
$\alpha^{(\partial_t)}_{-\tau}(\psi)$ are two different wavefunctions. 
This point of view will be useful shortly in a context where
time evolution makes no sense at all.\\
{\bf (2)} The found $SL(2,\bR)$ symmetry is only due to the shape of spectrum of $\sigma(H)$
which is $[0,+\infty)$. The absence of degeneracy implies that the representation is irreducible.
 From  a physical point of view,  invariance under the conformal group $SL(2,\bR)$  
could look very unexpected if $m>0$ since the theory admits the scale $m$. However, that scale 
does not affect the spectrum of $H$. In physical terms this is due to the gravitational
energy which is encompassed by $H$ itself since Rindler frame represents the 
spacetime experienced by an accelerated observer. \\
{\bf (3)} It is clear that the found $SL(2,\bR)$ symmetry can straightforwardly 
be extended to the free-field quantization by defining multi-particle operators generated by $H,C,D$.\\
{\bf (4)} Generators $iH_0,iD$, {\em differently} from $iH_0$ and $iC$, define a basis of the Lie algebra of
a subgroup of ${SL}(2,\bR)$, $SL^\triangle_+(2,\bR)$, made of real $2\times 2$ upper triangular 
matrices with unitary determinant and positive trace. 
(\ref{invariancetrue}) holds for $U_g$, $g\in SL^\triangle_+(2,\bR)$,
giving rise to another smaller symmetry of the system. The subgroup
generated by $iH_0$ trivially enjoys the same fact.\\

\noindent {\bf 2.5}.{\em Hidden and manifest symmetries}.
 A differentiable group of {\em geometric} symmetries of a classical Klein-Gordon field  in ${\bf R}$ (however everything we say can be extended 
to any globally hyperbolic spacetime along the procedures presented in \cite{Wald}) 
is defined as follows. Take a differentiable, locally-bijective, representation, $G\ni g \mapsto d_{g}$,
of a connected Lie group, $G$,
where $d_{g}: {\bf R} \to {\bf R}$
are isometries. The representation automatically induces a group of transformations
$\{\alpha_{g} \}_{g\in G}$ 
of scalar fields $f:{\bf R}\to \bR$ (or $\bC$),  with $\left(\alpha_{g}(f)\right)(x) := f\left(d_{g^{-1}}(x)\right)$.
As  $d_g$ are (orientation-preserving) 
isometries, $\alpha_g$ define {\em geometric symmetries} of the field in the sense that they transform elements of ${\cal S}$
into elements of ${\cal S}$ 
not affecting the symplectic form. If quantization is implemented,
solutions of the equation of motion in ${\cal S}$ 
are  associated with  one-particle quantum states through the decompositions 
(\ref{decomposition}),(\ref{decompositionm=0}).
Consider a group of  quantum symmetries in the sense of (\ref{invariancetrue}), 
described by a strongly-continuous representation of a Lie group $G$ in terms of
 unitary operators $\{U_g\}_{g\in G}$. 
In this picture,  the one-parameter unitary group generated by the Hamiltonian
 is assumed to be a subgroup of $\{U_g\}_{g\in G}$. If
 the symmetry ``does not depend on time'', i.e., $g(t)=g$ in (\ref{invariancetrue}),
that assumption can be dropped or, equivalently,
 the subgroup generated by the Hamiltonian can be considered in the center of $\{U_g\}_{g\in G}$ (i.e. 
it commutes with the other elements of the group).
 If $\{U_g\}_{g\in G}$ is related, 
by means of (\ref{decomposition}), (\ref{decompositionm=0}) and (\ref{decomposition+m=0}),
to a group of geometric symmetries $\{\alpha_g\}_{g\in G}$, that is
$\widetilde{\left(\alpha_{g}(\psi)\right)}_+ = U_g\tilde\psi_+$ for all $g\in G$ and $\psi\in {\cal S}$,
we say that the symmetry is {\em manifest}. Otherwise
we say that the symmetry is {\em hidden}.
In Rindler space, the quantum symmetry group $\{e^{i\tau H}\}_{\tau\in \bR}$ give rise to  manifest symmetry
 because it is associated
with the geometric group of symmetries $\{\alpha_\tau\}_{\tau\in \bR}$,
induced by the one-parameter group of isometries generated by
the  Killing vector $\partial_t$. The situation changes dramatically
if considering the whole $\widetilde{SL}(2,\bR)$ symmetry.
The space of Killing fields of ${\bf R}$ has a basis 
 $\partial_t,\partial_X,\partial_T$ with
$\{\partial_T,\partial_X\} = 0$,
$\{\partial_T,\partial_t\} = \partial_X$ and
$\{\partial_X,\partial_t\} = \partial_T$. 
($X$ and $T$ are the spatial and temporal coordinate of a Minkowski frame with
$\partial_t = X\partial_T +T\partial_X$.) 
It is trivially proven that no Killing field  $a\partial_t+b\partial_X+c\partial_T$ enjoys, with respect to $\partial_t$, the commutation rule that $D$
enjoys with respect to $H_0$ in (\ref{three}) so that no Killing field corresponding to $C$
makes sense. Summarizing, ${\bf R}$ cannot support isometry representations of $SL(2,\bR)$ (or
$\widetilde{SL}(2,\bR)$)
or the subgroup $SL^{\triangle}_+(2,\bR)$ generated by $H_0,D$. 
Hence the whole $SL(2,\bR)$ symmetry and that associated with $D$
are {\em hidden}. 

\section{Conformal field on the horizon.}

{\bf 3.1}.{\em Restriction to horizons}. In \cite{mopi02}, a similar analysis is performed for ${\bf AdS}_2$  spacetime
since the spectrum of the Hamiltonian of a particle has the same structure as that in
Rindler space.  However as a remarkable difference  $SL(2,\bR)$ is a {\em manifest} symmetry of a quantum particle 
moving in ${\bf AdS}_2$ because $SL(2,\bR)$ admits a representation in terms of ${\bf AdS}_2$ isometries.\\
Coming back to Rindler space viewed as a (open) submanifold of Minkowski spacetime, a natural question arises: 
``{\em Regardless the found $SL(2,\bR)$ symmetry
is hidden, does it  become manifest if one considers quantum field theory in an appropriate subregion of ${\bf R}\cup \partial {\bf R}$?}''\\
We shall see that investigation on this  natural question has several implications 
with holography because it naturally leads  to the formulation of a quantum field theory on the horizon 
which is algebraically and unitarily related with that formulated in the bulk,
but also it suggests that the symmetry of the theory is greater than $SL(2,\bR)$ it being described by
 a {\em Virasoro algebra}. \\
It is clear from 2.5 that the only region which could give a positive answer to the question is
the boundary $\partial {\bf R}$ of Rindler wedge, i.e. a {\em bifurcate Killing horizon}
made of three disjoint parts
${\bf F}\cup {\bf P}\cup {\bf S}$.
{\bf S} (a point in our $2D$ case) is the spacelike submanifold of Minkowski spacetime where 
the limit of the Killing field $\partial_t$ vanishes whereas  
the lightlike submanifold of Minkowski spacetime  ${\bf F}$
and ${\bf P}$ (the former in the causal future of the latter)
are the {\em future} and the {\em past} horizon respectively, where the limit of 
$\partial_t$ becomes lightlike but {\em not} vanishes. 
Since  the induced  metric on ${\bf F}\cup {\bf P}$ is degenerate,  the diffeomorphisms of ${\bf F}\cup {\bf P}$ 
can be viewed as isometries and the question about a possible {\em manifest} $SL(2,\bR)$ symmetry on the
horizon must be interpreted in that sense: The unitary representation has to be associated with a group of
diffeomorphisms induced by a Lie algebra of vector fields defined on the horizon.\\
To go on, let us investigate the limit of wavefunctions when the horizon is approached by Rindler-time evolution. 
To this end, consider the Rindler light coordinates (\ref{lc}).
 ${\bf F}$ is represented by $u\to +\infty$, $v\in \bR$ whereas 
${\bf P}$ is given by $v\to -\infty$, $u\in \bR$. 
 Coordinates $u,v$  actually cover the  Rindler space only, but, separately, the limit of $v$ is well defined on the lightlike submanifold 
${\bf F}$ and  
the limit of $u$ is well defined on the submanifold ${\bf P}$ and they define well-behaved global coordinate 
frames on these submanifolds. ({\it It holds in the $2D$ case. For greater dimension, $v$ (resp. $u$) 
 together with other ``transverse'' coordinates
defines global coordinates on ${\bf F}$ (resp. ${\bf P}$) as well.})
This can be proven by passing to Minkowski light coordinates $U:= T-X, V:= T+X$ which
 satisfy $V = e^{\ka v}$, $U = e^{-\ka u}$  in ${\bf R}$. So, from now $v$ and $u$ are also interpreted
  as coordinates on ${\bf F}$
 and ${\bf P}$ respectively. 
We have the following remarkable technical result (where,
if $a\in \bC$, ``$a + c.c.$'' means 
 ``$a\:\:+$ {\em complex conjugation of} $a$'')  \\

\noindent {\bf Proposition 3.1}. {\em Take $\psi \in {\cal S}$, with associated (Rindler) positive frequency parts
$\tilde\psi_+$ or $\tilde\psi^{(\alpha)}_+$
as in (\ref{decomposition}) and (\ref{decompositionm=0}) and
consider the evolution of $\psi$ in the whole Minkowski spacetime. 
In coordinate $v\in \bR$, the restriction of $\psi$ to ${\bf F}$ reads  respectively
for $m>0$ and $m=0$, 
\begin{eqnarray}
\psi(v)  =  \int_{\bR^+} \frac{e^{-iEv}}{\sqrt{4\pi E}}
 N_{m,\ka}(E)\tilde\psi_+(E)\: dE + \mbox{c.c.} \:\:,\:\:
 \psi(v)  = \int_{\bR^+} \frac{e^{-iEv}}{\sqrt{4\pi E}}
 \tilde\psi^{\mbox{\em\scriptsize (in)}}_+(E)\: dE + \mbox{c.c.}\label{limitH+}
 \end{eqnarray}
In coordinate $u\in \bR$, the restriction of $\psi$ to ${\bf P}$ reads  respectively
for $m>0$ and $m=0$,
\begin{eqnarray}
\psi(u)  = \int_{\bR^+}\frac{e^{-iEu}}{\sqrt{4\pi E}}
 \overline{N_{m,\ka}(E)}\tilde\psi_+(E)\: dE + \mbox{c.c.} \:\:,\:\:
 \psi(u)  =  \int_{\bR^+}\frac{e^{-iEu}}{\sqrt{4\pi E}}
 \tilde\psi^{\mbox{\em \scriptsize (out)}}_+(E)\: dE + \mbox{c.c.}
 \label{limitH-}\:.
 \end{eqnarray}
 The function $N_{m,\ka}$ (that is restricted to $\bR^+$ in (\ref{limitH+}) and (\ref{limitH-})) can be defined on the whole  $\bR$ as
 \begin{eqnarray}
  N_{m,\ka}(E):= e^{-i \frac{E}{\ka}\log \frac{m}{2\ka}}\: \mbox{\em sign}(E) \:\Gamma\left(\frac{iE}{\ka}\right) 
  \sqrt{\frac{E}{\ka\pi} 
 \sinh \frac{\pi E}{\ka}} \label{monster}\:.
 \end{eqnarray}
 It belongs to $C^{\infty}(\bR)$ and satisfies  $|N_{m,\ka}(E)|=1$ and  
 $\overline{N_{m,\ka}(E)} = -N_{m,\ka}(-E)$ for all $E\in \bR$.}\\

\noindent {\em Proof}. As $t=0$ is part of a Cauchy surface in Minkowski spacetime,
$\psi$ can uniquely be extended into a smooth solution of Klein Gordon equation in 
Minkowski spacetime, therefore it makes sense to consider its restriction to ${\bf P}$ or ${\bf F}$. As $\psi$ is smooth, those restrictions can be
computed by taking the limit of the function represented in light Rindler coordinates.
First consider the case $m=0$ and $u\to \infty$.
 One has $\Phi_{E}^{\mbox{\scriptsize (out)}}(t(u,v),y(u,v)) = 
\frac{e^{-iEu}}{\sqrt{4\pi E}}$ and $\Phi_{E}^{\mbox{\scriptsize (in)}}(t(u,v),y(u,v)) = 
\frac{e^{-iEv}}{\sqrt{4\pi E}}$. Insert these functions in (\ref{decompositionm=0}) and 
extend each integrations on the whole $\bR$ axis by defining $\tilde\psi^{(\alpha)}_+(E)=0$ for $E\leq 0$.
Using the properties of $\tilde\psi^{(\alpha)}_+$
stated in Proposition 2.1 before (\ref{decompositionm=0}) one sees that  $\psi(u,v)$
can be decomposed as a sum of two functions (one in the variable $u$ and the other in the variable $v$) 
which are the real part of the Fourier transform of a couple of $L^1(\bR)$ functions. 
Taking the limit
$u\to +\infty$ the function containing only modes {\em out} vanishes as a consequence of Rieman-Lebesgue lemma and 
(\ref{limitH+}) with $m=0$ arises. The case $m=0$ and $v\to -\infty$ is strongly analogous.
The case $m>0$ is based on the following expansion \cite{grads} at $x\to 0$ with $\omega$ fixed in $\bR$
\begin{eqnarray}
K_{i\omega}(x) = \frac{i\pi e^{\pi\omega/2}}{2}\left(\frac{ix}{2}\right)^{i\omega} \frac{1+ O_\omega(x^2)}{\Gamma(1+i\omega) \sinh(\pi\omega)}
-\frac{i\pi e^{-\pi\omega/2}}{2}\left(\frac{ix}{2}\right)^{-i\omega} 
\frac{1+ O'_\omega(x^2)}{\Gamma(1-i\omega) \sinh(\pi\omega)}\:,
\end{eqnarray}
where, for $\omega$ fixed, $|O_{\omega}(x^2)|\leq C_\omega|x|^2$ and $|O'_{\omega}(x^2)|\leq C'_\omega|x|^2$ for some
real finite $C_\omega,C'_\omega$, whereas 
for $x$ fixed in $\bR$, $\omega \mapsto |O_{\omega}(x^2)|$ and $\omega \mapsto |O'_{\omega}(x^2)|$ are bounded.
Inserting the expansion above in the expression (\ref{modes}) and taking the limit as $u\to \infty$
in (\ref{decomposition}), Riemann-Lebesgue's
lemma together with some trivial properties of $\Gamma$ function (\cite{grads}) produces  (\ref{limitH+})
for $m>0$. The function (\ref{monster})  is nothing but sign$(E)$ $e^{-iE(\log(m/2\ka))/\ka}\Gamma\left(\frac{iE}{\ka}\right) 
\left| \Gamma\left(\frac{iE}{\ka}\right) \right|^{-1}$ \cite{grads} and so $|N_{m,\ka}(E)|=1$ for $E\neq 0$
is trivially true.
$\Gamma(ix)$ is smooth along the real axis with a simple pole 
in $x=0$ that is canceled out by the zero of $\mbox{sign}(x)\sqrt{x\sinh x}$ that is smooth in the whole $\bR$. 
Thus
  $N_{m,\ka}\in C^{\infty}((-\infty,+\infty))$. $|N_{m,\ka}(E)|=1$ for $E= 0$ is trivially valid by continuity.
$\overline{N_{m,\ka}(E)} = -N_{m,\ka}(-E)$ is a straightforward consequence of $\overline{\Gamma(ix)}=\Gamma(-ix)$
for $x\in \bR$.
The  case $v\to -\infty$ can be proven similarly. 
$\Box$\\

\noindent From a pure mathematical point of view  
Proposition 3.1 shows that a solution in ${\cal S}$ of the massive Klein-Gordon equation in $2D$ Rindler space are completely 
determined  by its values on  {\em either} the  future {\em or} the past horizon, whereas, in the massless case,
a solution ${\cal S}$ is completely determined  by its values on {\em both} the future {\em and} the past horizon (see Chap. 5 of \cite{Wald} and  references quoted therein  for general results on these topics also in more general spacetimes).\\
As $|N_{m,\ka}(E)|=1$ we can write $N_{m,\ka}(E)= e^{i\rho_{m,\ka}(E)}$ where the phase $\rho_{m,\ka}(E)$
is real-valued.
The restriction of $\psi$ to the horizon ${\bf F}$ (the other case is analogous)
depends from the mass of the field through the phase $\rho_{m,\ka}$ only. As a trivial result 
we see that two (free scalar QFT) theories in ${\bf R}$ with different strictly-positive masses $m\neq m'$ and 
 Rindler vacua $\Psi_m,\Psi_{m'}$
turn out to be  unitarily equivalent.  This is due to the 
unitary operator $U : {\gF}({\cal H}_{m}) \to {\gF}({\cal H}_{m'})$ naturally defined by the requirement 
$U\Psi_{m} = \Psi_{m'}$ and induced by  the scalar-product-preserving map between positive frequency wavefunctions 
$$\psi_{+} \mapsto \psi'_{+} \:\:,\:\: \mbox{with} \:\:\: \psi'_{+}(E) 
:= e^{+i(\rho_{m,\ka}(E)-\rho_{m',\ka}(E))} \psi_{+}(E)\:\:\:\: \mbox{for all $E\geq 0$}$$
where $\psi_{+}\in {\cal H}_{m}$ and $\psi'_{+}\in {\cal H}_{m'}$.
Similarly, each theory is unitarily equivalent to the massless theory built up using only {\em in} modes. 
Avoiding any choice for the mass, one is naturally lead to consider the class of the ``fields defined on the
horizon'' 
\begin{eqnarray}
\psi(v)  =  \int_{\bR^+} \frac{e^{-iEv}}{\sqrt{4\pi E}}
 \tilde\psi_+(E)\: dE + \int_{\bR^+} \frac{e^{+iEv}}{\sqrt{4\pi E}}
 \overline{\tilde\psi_+(E)}\: dE \label{fieldonhorizon}
 \end{eqnarray}
as the object which makes possibles all those crossed unitary identifications and exists 
independently from the quantum fields defined in the bulk ${\bf R}$ with their own masses. 
We want to try to consider this object as a {\em quantum} field in a sense we 
go to specify. \\

\noindent {\bf 3.2}.{\em Local quantum field theory on {\bf F} and {\bf P}}. Following the procedure presented in 2.2 and 2.3 we want to show 
that it is possible to define a local quantum field theory on ${\bf F}$ which matches with that defined in the bulk.
(That idea is not new in the literature and it has been used in important works as that by Sewell \cite{Se82}).
First of all define the space of ``wavefunctions'' ${\cal S}(\bF)$. A suitable definition which will be useful
later,  is the following:  ${\cal S}(\bF)$ is the space $\mathsf{S}(\bR;\bR)$
of the real-valued Schwartz' functions on $\bR$
where $\bR$ is identified with ${\bf F}$ itself by means of the coordinate $v$. 
Actually the name ``wavefunction'' is not appropriate because there is no wave equation 
to fulfill in our context. 
As a consequence the correct point of view to interpret the formalism is the  Heisenberg's picture. 
 ${\cal S}(\bF)$  has a natural nondegenerate symplectic form which is {\em invariant under 
the diffeomorphisms
 of ${\bf F}$ which preserve its orientation}:
\begin{eqnarray} \Omega_{{\bf F}}({\varphi},\varphi') := \int_{{\bf F}} \varphi'  d{\varphi}
     - {\varphi} d\varphi' 
\label{scalarproductH}\:.\end{eqnarray}  
To define the one-particle Hilbert space, consider the modes
\begin{eqnarray}
F_{E}(v) := \frac{e^{-iEv}}{{\sqrt{4\pi E}}}\label{modesmH+}\:\:\:\:\:\mbox{with $E\in \bR^+$}
\:.\label{Fmodes}
\end{eqnarray} 
Analogous definitions can be given with analogous notations for the past  horizon ${\bf P}$
using modes as in (\ref{Fmodes}) with $-iEv$ replaced for $-iEu$.
The following pair of propositions can be simply proven using Fourier transform theory for 
Schwartz' functions.\\

\noindent {\bf Proposition 3.2. (Completeness of modes)}.
{\em 
If $\varphi$ belongs to ${\cal S}(\bF)$, the function   
\begin{eqnarray}
\bR^+\ni E \mapsto \tilde\varphi_{+}(E):= -i\Omega_{{\bf F}}\left(\overline{F_{E}},\psi\right)
\label{decompositionH+}\:. 
\end{eqnarray}
satisfies  $\tilde\varphi_{+}(E)=  \sqrt{E}g(E)$, 
$\overline{\tilde\varphi_{+}(E)}=  \sqrt{E}g(-E)$, where $g\in \mathsf{S}(\bR,\bR)$.
 Moreover, for $v\in \bR$ (\ref{fieldonhorizon}),
\begin{eqnarray}
\varphi(v) = \int_{0}^{+\infty}  F_{E}(v) \tilde\varphi_+(E)\: dE +
\int_{0}^{+\infty} \overline{F_{E}(v)\tilde\varphi_+(E)}\: dE\:. \label{decomposition2H+}
\end{eqnarray}
Similar results hold replacing ${\bf F}$ for ${\bf P}$ everywhere.} \\

\noindent {\bf Proposition 3.3. (Associated Hilbert spaces)}. {\em 
If $\varphi$ belongs to either ${\cal S}(\bF)$,
define the one-to-one associated {\em
positive-frequency 
wavefunction} 
 \begin{eqnarray}
\varphi_{+}(v) := \int_{0}^{+\infty} \: F_{E}(v) \tilde\varphi_{+}(E)\: dE  
\label{decomposition+H+}\:.
\end{eqnarray}
With that definition and for $\varphi_1,\varphi_2$ in ${\cal S}(\bF)$,  
it results $\Omega_{{\bf F}}({\varphi_{1+}},\varphi_{2+})=0$ whereas
\begin{eqnarray}
\langle \varphi_{1+},\varphi_{2+} \rangle_{{\bf F}} :=  -i\Omega_{{\bf F}}(\overline{\varphi_{1+}},\varphi_{2+}) \label{hermiteH+} \:,
\end{eqnarray}
is well-defined and
\begin{eqnarray}
\langle \varphi_{1+},\varphi_{2+} \rangle_{{\bf F}} = \int_{0}^{+\infty} \overline{\tilde\varphi_{1+}(E)} \tilde\varphi_{2+}(E) \: dE\:.
\end{eqnarray} The {\em one-particle Hilbert} space ${\cal H}_{\bf F}$
 is defined as the Hilbert completion 
of the space of finite complex linear combinations of functions $\varphi_+$, for all 
$\varphi$ in ${\cal S}(\bF)$, 
equipped with the extension of the scalar product (\ref{hermiteH+}) to complex linear combinations of arguments. It results
${\cal H}_{\bf F} \cong L^2(\bR^+,dE)$.
Similar results and definitions hold replacing ${\bf F}$ for ${\bf P}$ everywhere.}\\

\noindent {\bf Definition 3.1}. {\bf (Quantum field operators on horizons)}.
{\em Consider the symmetrized Fock space $\gF_{\bf F}({\cal H}_{{\bf F}})$
with vacuum state $\Psi_{\bf F}$ and scalar product $\langle\:,\rangle_{\bf F}$.
The {\em quantum field operator on} $\bf F$, $\Omega_{{\bf F}}(\:.,\hat \phi_{\bf F})$ is the 
symmetric-operator valued function
\begin{eqnarray}
 \varphi \mapsto 
\Omega(\varphi, \hat\phi_{\bf F})
 &:=& ia_{\bf F}(\overline{{\varphi}_+}) -ia_{\bf F}^{\dagger}(\varphi_+) \label{PIF}\:, 
\end{eqnarray}
 where $\varphi\in {\cal S}(\bF)$. $a_{\bf F}(\overline{\varphi_+})$ and $a_{\bf F}^{\dagger}(\varphi_+)$
 respectively denote the annihilation and construction operator associated with the one-particle state $\varphi_+$
which are defined in the dense invariant subspace $\gF_{0{\bf F}}$
spanned by all states containing a finite,  arbitrarily large, number of particles with states 
given by positive-frequency  wavefunctions. An analogous definition is given replacing ${\bf F}$ for ${\bf P}$ everywhere.}\\

\noindent  Operators $\Omega_{\bf F}(\varphi, \hat\phi_{\bf F})$
and $\Omega_{\bf P}(\varphi, \hat\phi_{\bf P})$
 are essentially self-adjoint on $\gF_{0{\bf F}}$
 and $\gF_{0{\bf P}}$ respectively
 since their elements are analytic vectors.\\ 
 We want to define an analogous procedure to that in the bulk (see (\ref{fsmeared})) for smearing field operators by means of ``functions''
instead of  ``wavefunctions''. The issue is however relevant because it  permits to introduce the analogue $E_{\bf F}$ of the causal propagator $E$
in spite of having no equation of motion in ${\bf F}$.
The idea is that something like (\ref{psif}) should work also in our context. 
A clear difficulty is that  there is no a diffeomorphism invariant measure which can be used in the
analogue of (\ref{psif}) in place of $d\mu_g$. On the other hand  integration of $k$-forms is 
diffeomorphism invariant on (oriented manifolds). Therefore we aspect that the space of ``functions''
used to smear quantum fields should properly be a space of $1$-forms rather than functions.
To go on we notice that {\em a posteriori} $E_{\bf F}$ has to fulfill something like $[\hat\phi_{\bf F}(v), \hat \phi_{\bf F}(v')] = -iE_{\bf F}(v,v')$.
By a formal but straightforward computation which uses $[a_E,a^\dagger_E] = \delta(E-E')$ and the analogue of (\ref{naive}) 
with $\Phi_E$ replaced for $F_E$, one finds that 
$i[\hat\phi_{\bf F}(v), \hat \phi_{\bf F}(v')] = \frac{1}{4}\mbox{sign}\: (v-v')$. This $v$-parametrized distribution
actually defines a well-behaved transformation from the space 
of  exact (smooth) $1$-forms in ${\bf F}$ with compact support
to the space of smooth functions on ${\bf F}$. As the functions $f\in {\cal S}(\bF)$ vanish (with all of  their derivatives) 
as $v\to \infty$,
if $\eta = df$
$$\int_{v'\in \bR} \mbox{sign}\: (v-v') \eta(v') = f(v) - (-f(v)) = 2f(v)\:.$$ 
In the following, ${\cal D}({\bf F})$ is the real space 
of the  $1$-forms $\eta=d\varphi$ such that $\varphi \in {\cal S}({\bf F})$.\\

\noindent {\bf Definition 3.2}. {\bf (Causal propagator and associated quantum field on horizons)}.
{\em With the given notations, the {\em causal propagator} in ${\bf F}$, is  the mapping
$E_{\bf F}:{\cal D}({\bf F})\to {\cal S}({\bf F})$
with
\begin{eqnarray} (E_{\bf F}\eta)(v) := \frac{1}{4} 
\int_{v'\in \bR} \mbox{sign}\: (v-v') \eta(v')
\:,\label{EF}\end{eqnarray}
and  the  quantum-field operator on {\bf F} {\em smeared  with forms $\eta$ 
of} ${\cal D}({\bf F})$   
is the mapping
\begin{eqnarray}
\eta \mapsto \hat\phi_{\bf F}(\eta) := \Omega_{\bf F}(E_{\bf F}\eta, \hat\phi_{\bf F})\:.\label{fFsmeared}
\end{eqnarray}
Analogous definitions are given replacing ${\bf F}$ for ${\bf P}$ and $v$ for $u$ everywhere.}\\

\noindent 
The given definitions are good generalizations of the analogous tools in usual quantum field theory 
 (see (\ref{psif}) and (\ref{locality}) in particular)
as stated in the following pair of propositions whose proof is trivial. The second item in Proposition 3.5 shows that the theory enjoys {\em locality} in a the same way as in Sewell's approach
\cite{Se82}. \\

\noindent {\bf Proposition 3.4.}
{\em If $\varphi \in {\cal S}(\bF)$, 
$\omega = 2 d\varphi$ is the  unique element of ${\cal D}({\bf F})$
such that $\varphi = E_{\bf F}(\omega)$. Moreover,
if $\eta,\omega \in 
{\cal D}({\bf F})$
\begin{eqnarray}
\int_{\bf F} \varphi \eta   = \Omega_{\bf F}(E_{\bf F}\eta,\varphi) \:\: \mbox{and}\:\:
\int_{\bf F} (E_{\bf F}\omega) \eta = \Omega_{\bf F}(E_{\bf F}\eta,E_{\bf F}\omega)
\label{psifF}\:.
\end{eqnarray}
An analogous result holds replacing ${\bf F}$ for ${\bf P}$ everywhere.}\\

\noindent {\bf Proposition 3.5.}
{\em If $\varphi \in {\cal S}({\bf F})$
and $\eta,\omega \in 
{\cal D}({\bf F})$
\begin{eqnarray}
\mbox{$[$}\hat\phi_{\bf F}(\eta),\hat\phi_{\bf F}(\omega)\mbox{$]$}
 &=& -iE(\eta,\omega):= -i \int_{\bf F} (E_{\bf F}\eta)\omega \:. \label{localityF}
\end{eqnarray}
In particular the following locality property holds true $$[\hat\phi_{\bf F}(\eta),\hat\phi_{\bf F}(\omega)]=0\:\:\:\:\:\: \mbox{if $\mbox{supp}\: \eta \cap \mbox{supp}\: \omega =\emptyset$.}$$
An analogous result holds replacing ${\bf F}$ for ${\bf P}$ everywhere.}\\

\noindent {\bf 3.3}.{\em The algebraic approach}. To state holographic theorems
it is necessary to re-formulate quantum field theory in an algebraic approach either in the bulk 
and on the horizon.  In globally hyperbolic spacetimes, 
linear QFT can be formulated independently from a preferred vacuum state and Fock representation.
It is worthwhile stressing that \cite{Wald} physics implies the existence of meaningful 
quantum states
which cannot be represented in the same Hilbert (Fock) representation of the algebra of observables. In this sense the algebraic
approach is more fundamental than the usual Fock approach in QFT in curved spacetime.
 Let us summarize the procedure in  
${\bf R}$ which, at least for $m>0$, could be replaced by any
globally hyperbolic spacetime.
The basic tool is an abstract  $*$-algebra, ${\cal A}_{\bf R}$,  
made of  the linear combinations of products of 
 formal field operators
$\phi(f),\phi(f)^*$ ($f\in {\cal D}({\bf R};\bC)$) and the unit $I$,
 which enjoy the same properties of operators $\hat\phi(f),\hat\phi(f)^\dagger$
(and the identity operator $I$).
 From a physical point of view, the Hermitian elements of ${\cal A}_{\bf R}$ 
represent the {\em local observables} of the free-field theory.
For $m>0$, the required properties are:\\
(1) $\phi(f)^*= \phi(\overline{f})$ for all $f\in {\cal D}({\bf R};\bC)$,\\ (2) $\phi(af + bg) = a\phi(f) + b\phi(g)$
for all $f,g \in {\cal D}({\bf R};\bC)$, $a,b\in \bC$, \\ (3) $[\phi(f), \phi(g)] = -iE(f,g)I$
for all $f,g \in {\cal D}({\bf R};\bC)$,
and\\  (4) $\phi(f)=0$ if (and only if) $f=Kh$ for some $h\in {\cal D}({\bf R};\bC)$.\\ 
${\cal A}_{\bf R}$ is rigorously realized as follows. Consider the complex unital algebra ${\cal A}_{0{\bf R}}$,
freely generated by the unit $I$, and abstract objects $\phi(f)$ and $\phi(f)^*$ for all  
$f\in {\cal D}({\bf R};\bC)$.  The involution $^*$ on  ${\cal A}_{0{\bf R}}$
is the unique antilinear involutive function $^*:{\cal A}_{0{\bf R}}\to {\cal A}_{0{\bf R}}$
such that $I^*=I$, $(\phi(f))^*= \phi(f)^*$. 
Let ${\cal I}\subset {\cal A}_{\bf R}$  be the double-side ideal 
whose elements are linear combinations of products containing at least one of the following factors
 $\phi(f)^*-\phi(\overline{f})$, $\phi(af + bg) - a\phi(f) - b\phi(g)$,  $[\phi(f), \phi(g)] +iE(f,g)I$,
and  $\phi(Kg)$  for  $f,g \in {\cal D}({\bf R};\bC)$, $a,b\in \bC$.
 ${\cal A}_{\bf R}$
 is defined as the  space of equivalence classes with respect to the equivalence relation
in ${\cal A}_{0{\bf R}}$, $A \sim B$ iff $A-B\in {\cal I}$ and ${\cal A}_{\bf R}$ is equipped with the 
$*$-algebra structure induced by ${\cal A}_{0{\bf R}}$ through $\sim$. 
If, with little misuse of notation, $\phi(f)$  and $I$ respectively  denote the classes $[\phi(f)]$ and $[I]\in {\cal A}_{\bf R}$, the 
properties (1),(2),(3), (4) are fulfilled.\\
If $m=0$, there are two relevant algebras ${\cal A}^{(in)}_{\bf R}$  and ${\cal A}^{(out)}_{\bf R}$.
${\cal A}^{(in)}_{\bf R}$ is the unital $*$-algebra  generated by $I$, 
${\phi}_{\mbox{\scriptsize in}}(f)$ and ${\phi}_{\mbox{\scriptsize in}}(f)^*$ for every $f\in {\cal D}({\bf R}, \bC)$
whereas  ${\cal A}^{(out)}_{\bf R}$ is the unital $*$-algebra  generated by $I$, 
${\phi}_{\mbox{\scriptsize out}}(f)$ and ${\phi}_{\mbox{\scriptsize out}}(f)^*$ for every $f\in {\cal D}({\bf R}, \bC)$.
By definition these algebras satisfy the constraints (1),(2),(3) and (4) with the difference that, in (3), $E$ must be replaced for
$E_{\mbox{\scriptsize in}}$ or $E_{\mbox{\scriptsize out}}$ respectively and, in (4), $K$ must be replaced by $\partial_u$ or 
$\partial_v$ respectively. The rigorous definitions can be given similarly to the case $m>0$, by starting from freely generated algebras 
and passing to quotient algebras.
We recall  that if  ${\cal A}$, ${\cal B}$
are $*$-algebras with field $\bC$ and units $I_{\cal A}$, $I_{\cal B}$, ${\cal A}\otimes {\cal B}$ 
denotes (see p.143 of \cite{algebre}) the $*$-algebra whose associated vector-space structure is the tensor product ${\cal A}\otimes {\cal B}$, 
the  unit is  $I:= I_{\cal A}\otimes I_{\cal B}$,
the involution and the algebra product are respectively  given by $(\sum_k A_k\otimes B_k)^*:=\sum_k  A_k^*\otimes B_k^*$ and  
$(\sum_k A_k\otimes B_k) (\sum_i A'_i\otimes B'_i):= \sum_{ki}A_kA'_i\otimes  B_kB'_i$
with obvious notation.
Assuming (\ref{decm=0}) as the definition of $\phi(f)$, the whole field algebra ${\cal A}_{\bf R}$ is {\em defined} as
${\cal A}_{\bf R}:={\cal A}^{(in)}_{\bf R}\otimes {\cal A}^{(out)}_{\bf R}$. 
That unital $*$-algebra  satisfies (1),(2),(3) and (4).\\
An {\em algebraic state} on a $*$-algebra  ${\cal A}$ with unit $I$, is a linear map $\mu : {\cal A} \to \bC$ that is
 normalized (i.e. $\mu(I)=1$) and positive
(i.e. $\mu(A^*A)\geq 0$ for $A\in {\cal A}$). The celebrated GNS theorem (e.g., see \cite{Wald}) states that
for every algebraic state $\mu$ on ${\cal A}$ there is a  triple 
$(\gH_\mu, \Pi_\mu, \Omega_\mu)$ such that  the following facts hold.
 $\gH_\mu$ is a Hilbert space, $\Pi_\mu$ is a $*$-algebra 
representation  of ${\cal A}$ in terms of operators on $\gH_\mu$ which are
defined on a dense invariant subspace 
${\gD}_\mu \subset \gH_\mu$ and such that $\Pi_\mu(A^*)= (\Pi_\mu(A))^\dagger\spa\rest_{{\gD}_\mu}$.
Finally ${\gD}_{\mu}$ is spanned by all the vectors $\Pi_\mu(A)\Omega_\mu$, $A\in {\cal A}$,
and $\mu(A)= \left\langle\Omega_\mu, \Pi_\mu(A)\Omega_\mu\right\rangle$ for all $A\in {\cal A}$, $\langle \:,\:\rangle$ 
denoting the scalar product in
$\gH_\mu$. If  $({\gH}'_\mu, \Pi'_\mu, \Omega'_\mu)$ 
is another similar triple associated with the same $\mu$, there is a unitary operator 
$U: \gH_\mu \to {\gH}'_\mu$  such that $\Omega'_\mu = U\Omega_\mu$ and $\Pi'_\mu = U\Pi_\mu$.
If ${\cal A}={\cal A}_{\bf F}$, by direct inspection one finds that  quantum field theory in ${\bf R}$ in the Fock-space $\gF({\cal H})$
with $\Psi_{\bf R}$ as vacuum state
coincides with that in a GNS representation
of ${\cal A}_{\bf R}$ associated with the ({\em quasifree} \cite{Wald}) algebraic state 
$\mu_{\bf R}$ completely determined by
$\mu_{\bf R}(\phi(f)\phi(g)):= \left\langle\Psi_{\bf R},
\hat \phi(f)\hat\phi(g) \Psi_{\bf R}\right\rangle$ via Wick expansion of symmetrized $n$-point functions. 
Moreover  it results ${\gD}_{\mu}={\gF}_0$. \\
 All the procedure can be used to give an algebraic approach for QFT on ${\bf F}$ (or {\bf P}): Define ${\cal D}({\bf F};\bC):=
{\cal D}({\bf F})+ i{\cal D}({\bf F})$
and define $\hat\phi_{\bf F}(\omega+i\eta):= \hat\phi_{\bf F}(\omega) + i \hat \phi_{\bf F}(\eta)$ when $\omega,\eta\in {\cal D}({\bf F})$. 
Finally, consider the abstract  $*$-algebra ${\cal A}_{\bf F}$ 
with unit $I$, 
 generated by $I$, $\phi_{\bf F}(\omega)$, $\phi_{\bf F}(\omega)^*$ for all
$\omega \in {\cal D}({\bf F};\bC)$,
such that,  for all $a,b\in \bC$ and  $\omega,\eta\in {\cal D}({\bf F};\bC)$:  \\
(1) $\phi_{\bf F}(\omega)^*= \phi_{\bf F}(\overline{\omega})$,\\ (2) $\phi_{\bf F}(a\omega + b\eta) = a\phi_{\bf F}(\omega) + b\phi_{\bf F}(\eta)$ and\\  
(3) $[\phi_{\bf F}(\omega), \phi_{\bf F}(\eta)] = -iE_{\bf F}(\omega,\eta)I$.\\
(The rigorous definition is given in terms of quotient algebras as usual.)
 From a physical point of view the (Hermitean) elements of ${\cal A}_{\bf F}$  represents the 
{\em (quasi) local observables} of the free-field theory on the future horizon.
By direct inspection one finds that quantum field theory in ${\bf F}$ in the Fock-space referred to the vacuum state $\Psi_{\bf F}$
coincides with that in a GNS representation of  $*$-algebra ${\cal A}_{\bf F}$ 
associated with  a the (quasifree) algebraic state 
$\mu_{\bf F}$ completely determined, via Wick expansion, by
$\mu_{\bf F}(\phi_{\bf F}(\eta)\phi_{\bf F}(\omega)):= \left\langle\Psi_{\bf F},
\hat \phi_{\bf F}(\eta)\hat\phi_{\bf F}(\omega) \Psi_{\bf F}\right\rangle_{\bf F}$
and ${\gD}_{\mu}={\gF}_{0{\bf F}}$.\\
Everything can similarly be stated for quantum filed theory on ${\bf P}$ with trivial changes
in notations.\\

\noindent {\bf 3.4}. {\em Two holographic theorems}. Here we  prove two {\em holographic} theorems for the observables of 
free fields, one in the algebraic approach and the latter 
in the Hilbert-space formulation under the choice of suitable vacuum states. 
The former theorem says that, in the massive case, there is a one-to-one transformation 
from the algebra of the fields in the bulk ${\cal A}_{\bf R}$ 
 -- that is the local observables of the free field in the bulk --
 to a subalgebra of fields on the future horizon ${\cal A}_{\bf F}$
-- that is the observables of the free field in the future horizon --.
The mapping preserves the structure of $*$-algebra and thus the two classes of obsevables can be identified completely
nomatter the value of the mass of the field in the bulk and the fact that there is no mass associated with the field on the horizon.
Remarkably, this identification does not requires any choice for reference vacuum states since it is given at a pure algebraic level. 
 To build up the said mapping,
 take any compactly supported function $f$ in the bulk, consider the generated wavefunction $\psi_f=E(f)$
(that is assumed to be defined in the whole Minkowski space), restrict $\psi_f$ on ${\bf F}$ obtaining a horizon wavefunction 
 $\varphi_f$ 
and associate 
with that function the unique form $\omega_f$ with $\varphi_f = E_{\bf F}\omega_f$. 
Finally define $\chi_{\bf F}(\phi(f)):=\phi_{\bf F}(\omega_f)$. Next step  is to extend 
 $\chi_{\bf F}$ to the whole algebra ${\cal A}_{\bf R}$ by requiring that 
 the $*$-algebra structure is preserved 
that is, $I$ is mapped in $I$, $\phi(f)^*$ is mapped into $\chi_{\bf F}(\phi(f))^*$, products of fields $\phi(f)\phi(g)$
are mapped into  $\chi_{\bf F}(\phi(f))\chi_{\bf F}(\phi(g))$ and so on. In the massless case, the procedure is similar but one 
has to consider also the past evolution of wavefunctions toward the past  horizon ${\bf P}$.
The theorem says that the required extensions into  algebra homomorphisms actually exists, are uniquely determined and injective
so that the observable algebra in  the bulk can be seen as a observable subalgebra on the horizon.\\

\noindent {\bf Theorem 3.1}. ({\bf Algebraic holography}) {\em In a $2D$-Rindler space ${\bf R}$
viewed as immersed in a corresponding $2D$ Minkowski spacetime,
consider quantum field theory of a scalar field with mass $m\geq 0$ satisfying Klein-Gordon equation 
(\ref{KG}). Consider the algebra
${\cal A}_{\bf R}$ (including 
${\cal A}^{\mbox{\scriptsize (out)}}_{\bf R}$, ${\cal A}^{\mbox{\scriptsize (in)}}_{\bf R}$ if $m=0$) 
of local observables in the bulk 
and the algebras ${\cal A}_{\bf F}$ and  ${\cal A}_{\bf P}$ of the observables on the horizons ${\bf F}$ and ${\bf P}$. 
The following statements hold.\\
{\bf (a)} If $m>0$,
 there is a  unital-$*$-algebra  homomorphism
$\chi_{\bf F}: {\cal A}_{\bf R} \to {\cal A}_{\bf F}$ uniquely determined by
\begin{eqnarray}
\chi_{\bf F}: \phi(f) \mapsto \phi_{\bf F}(\omega_f) \:\:\:\: \mbox{with $\omega_f := {2}d ((Ef)\spa\rest_{\bf F})$
for all 
$f\in {\cal D}({\bf R})$}  
\label{TH1m>0}\:,
\end{eqnarray}
$(Ef)\spa\rest_{\bf F}$ denoting the limit of $E(f)$ on ${\bf F}$.
$\chi_{\bf F}$ turns out to be  injective.\\
An analogous statement holds replacing ${\bf F}$ for ${\bf P}$.\\
{\bf (b)} If $m=0$, there are two  unital-$*$-algebra  homomorphisms
$\pi_{\bf P}: {\cal A}^{\mbox{\scriptsize (out)}}_{\bf R} \to {\cal A}_{\bf P}$  and 
$\pi_{\bf F}: {\cal A}^{\mbox{\scriptsize (in)}}_{\bf R} \to {\cal A}_{\bf F}$
uniquely determined by
\begin{align}
\pi_{\bf F} &: \phi_{\mbox{\scriptsize in}}(f) \mapsto \phi_{\bf F}(\eta_f)  
\:\:\:\:&\mbox{with $\omega_f := {2}d (E(f)\spa\rest_{\bf F})$ for all $f\in {\cal D}({\bf R})$}\:, \label{TH1m=01}\\
 \pi_{\bf P} &: \phi_{\mbox{\scriptsize out}}(f)  \mapsto \phi_{\bf P}(\omega_f) 
\:\:\:\: &\mbox{with $\eta_f := {2}d (E(f)\spa\rest_{\bf P})$ for all $f\in {\cal D}({\bf R})$}
\label{TH1m=02}\:.
\end{align}
$\pi_{\bf F}$ and $\pi_{\bf P}$ turn out to be  injective.\\
{\bf (c)} $\pi_{\bf F}({\cal A}^{\mbox{\scriptsize (in)}}_{\bf R})\subset {\cal A}_{{\bf F}}$ 
is  the subalgebra generated by $I$ and abstracts field operators smeared by the elements of 
${\cal D}({\bf F},\bC)$ with compact support.
The analogous statement holds for $\pi_{\bf P}({\cal A}^{\mbox{\scriptsize (out)}}_{\bf R})$.}\\

\noindent {\em Proof}. 
{\bf (a)}  The uniqueness of the homomorphism is trivially proven  by noticing that the elements of 
${\cal A}_{\bf R}$ are of the form $A= aI + \sum_kb_k\phi(f_k) + \sum_hc_h\phi(g_h)^* +\sum_{ls}d_{ls}\phi(h_l)\phi(p_s)+...$
where the overall summation as well as every partial summation is finite. As $\chi_{\bf F}$ is a 
homomorphism, $\chi_{\bf F}(A) = aI + \sum_kb_k\chi_{\bf F}(\phi(f_k)) + \sum_hc_h\chi_{\bf F}(\phi(g_h))^* +
\sum_{ls}d_{ls}\chi_{\bf F}(\phi(h_l))\chi_{\bf F}(\phi(p_s))+...$ 
Moreover $\chi_{\bf F}(\phi(h))= \chi_{\bf F}(\phi(Re\: h))+ i\chi_{\bf F}(\phi(Im\: h))$
and thus the values $\chi_{\bf F}(\phi(f))$ with $h$ real
determine the homomorphism provided it exists. 
Let us prove the existence of the homomorphism.
Take $f\in {\cal D}({\bf R})$ and consider $\psi_f=Ef$ and the associated function 
$\tilde{\psi_f}_+=\tilde{\psi_f}_+(E)$. It holds $\tilde{\psi_f}_+(E)= \sqrt{E}f(E)$
with $f\in {\mathsf S}(\bR,\bC)$ such that $\overline{f(E)} = -f(-E)$ as stated in Proposition 2.1 
and $N_{m,\ka}\in C^\infty(\bR)$ (with $|N_{m,\ka}(E)|=1$) and $\overline{N_{m,\ka}(E)} = -N_{m,\ka}(-E)$
as stated in Proposition 3.1. As a consequence $N_{m,\ka}(E)\tilde{\psi}_+(E) = \sqrt{E}h(E)$
where $\overline{h(E)} = h(-E)$ and $h\in {\mathsf S}(\bR,\bC)$.
Passing to the function $v\mapsto \psi_f(v)$ in Proposition 3.1 and using these results
one gets 
$$\psi_f(v)= \mbox{const.}\int_{0}^{+\infty}  e^{-iEv} h(E)dE + \mbox{c.c.}= \mbox{const.} \:\int_{-\infty}^{+\infty}  e^{-iEv} h(E)dE\:.$$
As $h$ belongs to Schwartz' space, $\psi_f$  belongs to the same  space because Fourier transform
preserves Schwartz' space. Moreover $\psi_f$ is real since $h(E) = \overline{h(-E)}$. We have found that  
$\psi_f\in {\cal S}(\bF)$ and thus $\omega_f:= {2}d\psi_f = {2}d [(Ef)\spa\rest_{\bf F}]$ 
is an element of ${\cal D}({\bf F})$. Using $f\in {\cal D}({\bf R},\bC)$ the result is preserved trivially
by the linear decomposition in real and imaginary part. Assume once again that $f\in {\cal D}({\bf R})$.  
Notice that $\omega_f$ contains the same information as 
$\psi_f$ because  $\psi_f(v) = 2\int_{-\infty}^{v} \omega_f$.
In turn $\psi_f$ determines the function $\tilde{\psi_f}_+$ which determines $Ef$. As we said in {2.3}, $Ef$ determines $f$ up to 
a term $Kh$ with $h\in {\cal D}({\bf R})$. We conclude that  $\omega_f=\omega_g$ if and only if $f=g + Kh$ with 
$h\in {\cal D}({\bf R})$. The same result arises for functions  $f,g\in {\cal D}({\bf R},\bC)$
by linearity and from the fact that $E$ transforms real functions into real functions.
We have found that there is a well-defined linear map ${\cal D}({\bf R},\bC)\ni f\mapsto 
\omega_f\in {\cal D}({\bf F},\bC)$ that transforms real functions in real forms and such that $\omega_f=\omega_g$ if and only if
$f=g + Kh$.
Now we define $\chi_{0\bf F}(\phi(f))=\phi_{\bf F}(\omega_f)$ and $\chi_{\bf F}(I)=I$ and $\chi_{\bf F}(\phi(f)^*)=
\phi_{\bf F}(\omega_f)^*$.
That map extends straightforwardly from the 
$*$-algebra ${\cal A}_{0{\bf R}}$ freely generated by $I$, $\phi(f)$, $\phi(f)^*$ (with involution uniquely determined as said in 
2.4) to the analogous free $*$-algebra 
on ${\bf F}$ giving rise to a $*$-algebra homomorphism $\chi_{0\bf F}$. However it is not injective since it results 
$\chi_{0\bf F}(\phi(f))=\chi_{0\bf F}(\phi(g))$
whenever $f=g+Kh$, and more generally injectivity failure arises for any pair of elements of the algebra which are different 
to each other because of the presence of
factors $\phi(f)$ and $\phi(g)$ with $f-g=Kh$. The injectivity is however restored if we take the quotient $*$-algebra 
${\cal A}_{1{\bf R}}$ 
in ${\cal A}_{0{\bf R}}$ with respect to the both-side ideal containing linear combinations of products with at least 
one factor $\phi(Kf)$ or $\phi(Kf)^*$ for any $f\in {\cal D}({\bf F},\bC)$ and redefine the injective 
map $\chi_{1\bf F}:{\cal A}_{1{\bf R}} \to {\cal A}_{0{\bf F}}$ as 
the map induced by $\chi_{0\bf F}$ through the canonical projection of ${\cal A}_{0{\bf R}}$ onto  ${\cal A}_{1{\bf R}}$.
By construction $\chi_{1\bf F}$ is an injective $*$-algebra isomorphism. In this context and from now on, 
it is convenient to think the objects $\phi(f)$ as smeared by the equivalence class $[f]$ rather than $f$ itself, where
$[f]$ belong to the complex vector space obtained by taking the quotient of ${\cal D}({\bf R},\bC)$ 
with respect to the subspace $K{\cal D}({\bf R},\bC)$. The map $[f] \mapsto \omega_f$ is a well-defined injective
vector space isomorphism that preserves the complex conjugation.
To conclude we have to extract the algebras ${\cal A}_{\bf R}$ and ${\cal A}_{\bf F}$ by the procedure outlined in 2.4,
 based on the projection on suitable quotient spaces,
and prove that the $*$-homeomorphism $\chi_{1{\bf F}}$ induces a $*$-homeomorphism $\chi_{\bf F}: {\cal A}_{\bf R}\to 
{\cal A}_{\bf F}$. To this end we have to consider the double-side ideal 
 ${\cal I}\subset {\cal A}_{1\bf R}$  
whose elements are linear combinations of products containing at least one of the following factors
 $\phi(f)^*-\phi(\overline{f})$, $\phi(af + bg) - a\phi(f) - b\phi(g)$,  $[\phi(f), \phi(g)] +iE(f,g)I$,
for  $f,g \in {\cal D}({\bf R};\bC)$, $a,b\in \bC$.
 ${\cal A}_{\bf R}$
 is defined as the  space of equivalence classes with respect to the equivalence relation
in ${\cal A}_{1{\bf R}}$, $A \sim_{\cal I} B$ iff $A-B\in {\cal I}$ and ${\cal A}_{\bf R}$ is equipped with the 
$*$-algebra structure induced by ${\cal A}_{1{\bf R}}$ through the canonical projection. 
The analogous procedure must be used for ${\cal A}_{1{\bf F}}$ with respect to 
an analogous ideal ${\cal I}_{\bf F}\subset {\cal A}_{1\bf F}$ in order to produce ${\cal A}_{{\bf F}}$.
Then the injective $*$-homomorphism $\chi_{1{\bf F}}$ induces a injective $*$-homomorphism $\chi_{\bf F}: {\cal A}_{\bf R}\to 
{\cal A}_{\bf F}$ if the equivalence relations induced by ${\cal I}$ and ${\cal I}_{\bf F}$ are
 preserved by $\chi_{1{\bf F}}$ itself, i.e. $A\sim_{\cal I} B$ if and only if 
 $\chi_{1{\bf F}}(A) \sim_{{\cal I}_{\bf F}} \chi_{1{\bf F}}(B)$.
We leave the simple but tedious proof of this fact to the reader, proving the only non trivial point which concerns 
factors $[\phi(f), \phi(g)] +iE(f,g)I$. It is simply found that, among other trivially fulfilled conditions,
the preservation of the equivalence relation 
arises if $\chi_{1{\bf F}}([\phi(f), \phi(g)] +iE(f,g)I)= 
[\phi_{\bf F}(\omega_f), \phi_{\bf F}(\omega_g)] +iE_{\bf F}(\omega_f,\omega_g)I$. Which is equivalent to
$E(f,g)=E_{\bf F}(\omega_f,\omega_g)$. (Notice that, by the known properties of the causal propagator 
both sides are invariant under the addition of 
a term $Kh$  to $f$ or $g$.)  $E(f,g)=E_{\bf F}(\omega_f,\omega_g)$ is equivalent to, with obvious notations,
 $\Omega(\psi_f,\psi_g)=\Omega_{\bf F}(\varphi_f,\varphi_g)$. It is sufficient to prove that identity  for real $f,g$.
 By Propositions 2.2 and 3.3  one finds $-i\Omega(\psi_f,\psi_g)= \langle \psi_{f+}, \psi_{g+}\rangle -  
 \overline{\langle \psi_{f+}, \psi_{g+}\rangle}$ and $-i\Omega_{\bf F}(\varphi_f,\varphi_g)= 
 \langle \varphi_{f+}, \varphi_{g+}\rangle_{\bf F} -  
 \overline{\langle \varphi_{f+}, \varphi_{g+}\rangle_{\bf F}}$. Passing in energy representation, where the scalar product
 is simply that of $L^2(\bR^+,dE)$ in both spaces,
  $\psi_{f+}$ and $\psi_{g+}$ are represented by some $E\mapsto \tilde{\psi}_{f+}(E)$ and $E\mapsto\tilde{\psi}_{g+}(E)$
  respectively whereas, by Proposition 3.1, $\varphi_{f+}$ and $\varphi_{g+}$ are represented by 
  $E\mapsto N_{m,\ka}(E)\tilde{\psi}_{f+}(E)$ and $E\mapsto N_{m,\ka}(E)\tilde{\psi}_{g+}(E)$ respectively. Since $|N_{m,\ka}(E)|=1$
  it results $\langle \psi_{f+}, \psi_{g+}\rangle = \langle \varphi_{f+}, \varphi_{g+}\rangle_{\bf F}$ that entails 
 $\Omega(\psi_f,\psi_g)=\Omega_{\bf F}(\varphi_f,\varphi_g)$ and concludes the proof.
 {\bf (b)}  Following a proof very similar to that as in the case
(b) (but simpler since the phases $N_{m,\ka}$ disappear when one uses Proposition 3.1) one sees that 
${\cal A}^{\mbox{\scriptsize(in)}}_{\bf R}$ is isomorphic to ${\cal A}_{\bf F}$ 
under the unique extension, into a injective $*$-algebra-with-unit homomorphism, of the 
map $\pi_{\bf F}: {\phi}_{\mbox{\scriptsize in}}(f)\mapsto\phi_{\bf F}(\omega_f)$ with 
$\omega_f := {2}d (E_{\mbox{\scriptsize in}}(f))$ and this is equivalent to the thesis 
because  $E_{\mbox{\scriptsize in}}f =E(f)\spa\rest_{\bf F}$
since $E_{\mbox{\scriptsize out}}(f)\spa\rest_{\bf F}=0$ 
and $E_{\mbox{\scriptsize in}}(f)\spa\rest_{\bf F}= E_{\mbox{\scriptsize in}}f$,
for smooth compactly supported
$f$ defined in ${\bf R}$ (these facts are consequences of Proposition 3.1). 
The case of  ${\cal A}^{\mbox{\scriptsize (out)}}_{\bf R}$ is strongly analogous. 
{\bf (c)} Consider the case of $\pi_{\bf F}$ the other being analogous. If $f$ is smooth and compactly supported in ${\bf R}$, 
$E_{\mbox{\scriptsize in}}f$ is a compactly-supported function of $v$ and thus 
$\omega_f = {2}d (E(f)\spa\rest_{\bf F}) = {2}d (E_{\mbox{\scriptsize in}}f)$ 
is compactly supported on ${\bf F}$.
Conversely if $\omega =d\varphi \in {\cal D}({\bf F},\bC)$ is  compactly supported  on ${\bf F}$, $\varphi$ must 
be compactly supported and
$f(u,v) := 2\varphi(u) h(u)$ is smooth,  compactly supported in ${\bf R}$ for every smooth compactly supported function
$h: \bR\to \bR$  and $\omega = {2}d (E_{\mbox{\scriptsize in}}(f))$ if $\int_\bR h(u) du =1$.
$\Box$\\

\noindent {\bf Remarks}. {\bf (1)} We stress that QFT on the horizon is the same nomatter the value of the
mass of the filed in the bulk: Different choices of the mass determine different injective $*$-algebra homomorphisms
from the algebra in the bulk to  the {\em same} algebra of observables on the horizon.\\
{\bf (2)} There are strong differences between the cases $m>0$ and $m=0$. If $f$ is compactly supported in the bulk, 
the horizon restriction of $Ef$ is compactly supported if $m=0$ but that is not the case when $m>0$. 
For that reason we have defined ${\cal S}({\bf F})$ (and ${\cal D}({\bf F})$) as a space of  
rapidly decreasing functions ($1$-forms) rather than
a space of compactly supported functions ($1$-forms). Moreover, if $m>0$, ${\cal A}_{\bf R}$ is isomorphic to 
a subalgebra of ${\cal A}_{\bf F}$ (or equivalently ${\cal A}_{\bf P}$). Conversely, if $m=0$, 
${\cal A}_{\bf R}(={\cal A}^{\mbox{\scriptsize (in)}}_{\bf R}\otimes {\cal A}^{\mbox{\scriptsize (out)}}_{\bf R})$ is isomorphic to 
a subalgebra of ${\cal A}_{\bf F}\otimes{\cal A}_{\bf P}$ by means of the injective unital-$*$-algebra homomorphism 
$\pi_{\bf P}\otimes \pi_{\bf F}: {\cal A}_{\bf R} \to {\cal A}_{\bf P}\otimes
{\cal A}_{\bf F}$.\\
{\bf (3)}  The existence of the  $*$-homorphisms
$\chi_{\bf F}$ and $\pi_{{\bf F}/{\bf P}}$ implies that, for all $f,g\in {\cal D}({\bf F},\bC)$ or 
${\cal D}({\bf F},\bC)$ and respectively for $m>0$ or $m=0$, 
\begin{equation}
\mbox{$[$}\phi(f),\phi(g)\mbox{$]$} = \mbox{$[$}\phi_{\bf F}(\omega_f),\phi_{\bf F}(\omega_g)\mbox{$]$} \:\: \mbox{or}\:\:
\mbox{$[$}\phi_{\mbox{\scriptsize in/out}}(f),\phi_{\mbox{\scriptsize in/out}}(g)\mbox{$]$} = 
\mbox{$[$}\phi_{{\bf F}/{\bf P}}(\omega_f),\phi_{{\bf F}/{\bf P}}(\omega_g)\mbox{$]$}
\label{UNO}\:,
\end{equation}
as a consequence the {\em causal propagator and the symplectic forms are preserved too}.\\

\noindent The second theorem concerns the unitary implementation of the $*$-homomorphism given in 
Theorem 3.1. This theorem states that, if one realizes the algebras of observables ${\cal A}_{\bf R}$
and ${\cal A}_{\bf P}$, ${\cal A}_{\bf F}$ in terms of proper field operators in the Fock spaces 
constructed over respectively, the Rindler vacuum $\Psi_{\bf R}$ and $\Psi_{\bf P}$, $\Psi_{\bf F}$, then
the injective homomorphisms presented in Theorem 3.1 are implemented by unitary operators which
preserve the vacuum states. In other words, {\em with the said choice of the vacuum states and Fock representation of the algebras
of observables}, the theory in the bulk and that on the horizon are {\em unitarily equivalent}.
As an immediate consequence, it arises that the {\em vacuum expectation values} are preserved passing from the theory 
 in the bulk ${\bf R}$ to the theory on the horizon ${\bf F}$ (or ${\bf P}$).\\

\noindent {\bf Theorem 3.2}. ({\bf Unitary holography}). {\em In the same hypotheses as in Theorem 3.1, consider 
the realization of the  algebra of the local observables of the bulk ${\cal A}_{\bf R}$, 
in the Fock space $\gF({\cal H})$ with Rindler vacuum $\Psi_{\bf R}$ 
 ($= \Psi^{\mbox{\scriptsize (out)}}_{\bf R}\otimes \Psi^{\mbox{\scriptsize (in)}}_{\bf R}$ if $m=0$)
 and the realizations of the algebras of observables of the horizons ${\cal A}_{\bf P}$, ${\cal A}_{\bf F}$
 in the Fock spaces $\gF({\cal H}_{{\bf P}})$, $\gF({\cal H}_{{\bf F}})$ of Definition 3.1
 with horizon vacua $\Psi_{\bf P}$,$\Psi_{\bf F}$.  With these realizations, the homomorphisms $\chi_{{\bf P}/{\bf F}}$ and $\pi_{_{{\bf P}/{\bf F}}}$
 can be implemented by unitary transformations which preserves the vacuum states. More precisely:\\
 {\bf (a)} If $m>0$,  the map that associates a positive frequency wavefunction ${\psi}_+$ in Rindler space with 
the element of ${\cal H}_{\bf F}\cong L^2(\bR^+,dE)$, $\phi: E \mapsto  M_{m,\ka}(E)\tilde{\psi}_+(E)$  extends
 into the unitary operator $U_{\bf F}: \gF({\cal H}) \to \gF({\cal H}_{{\bf F}})$ such that
 \begin{eqnarray} U_{\bf F} \Psi_{\bf R} &=& \Psi_{\bf F}\\
\chi_{\bf F}(\hat A) &=& U_{\bf F}\hat A U_{\bf F}^{-1}\:\:\:\:\:\:\:\:\:\: \:\:\:\:\:\:\:\:\:\mbox{for all $\hat A\in {\cal A}_{\bf R}$,}
 \end{eqnarray}
 The analogous statement holds replacing ${\bf F}$ for ${\bf P}$.\\
 {\bf (b)} If $m=0$, the maps which associate  positive frequency wavefunctions 
  ${\psi}^{\mbox{\scriptsize(in)}}_+$ and ${\psi}^{\mbox{\scriptsize(out)}}_+$
in Rindler space  with respectively elements of ${\cal H}_{\bf F}\cong L^2(\bR^+,dE)$ and ${\cal H}_{\bf P}\cong L^2(\bR^+,dE)$,
 $\phi^{\mbox{\scriptsize(in)}}: E\mapsto \tilde{\psi}^{\mbox{\scriptsize(in)}}_+(E)$
 and $\phi^{\mbox{\scriptsize(out)}}: E\mapsto \tilde{\psi}^{\mbox{\scriptsize(out)}}_+(E)$,
  extend into unitary operators $V_{\bf F}: \gF({\cal H}_{\mbox{\scriptsize (in)}}) \to \gF({\cal H}_{{\bf F}})$
 and $V_{\bf P}: \gF({\cal H}_{\mbox{\scriptsize (out)}}) \to \gF({\cal H}_{{\bf P}})$, 
 such that 
  \begin{eqnarray} V_{\bf F} \Psi^{\mbox{\scriptsize (in)}}_{\bf R} &=& \Psi_{\bf F}\:\:,\:\:\:\:\:\:\:  
   V_{\bf P} \Psi^{\mbox{\scriptsize (out)}}_{\bf R} = \Psi_{\bf P}\:, \\
  \pi_{\bf F}(\hat A) &=& V_{\bf F}\:\hat A \:V_{\bf F}^{-1}\:\:\:\:\:\:\:\:\:\:\:\:\:\:\:\:\:\:\:\:\:\:\:\: 
  \mbox{for all $\hat A\in {\cal A}^{\mbox{\scriptsize (in)}}_{\bf R}$,}\\
  \pi_{\bf P}(\hat A) &=&  V_{\bf P}\:\hat A \:V_{\bf P}^{-1}\:\:\:\: \:\:\:\:\:\:\:\:\:\:\:\:\:\:\:\:\:\:\:\:
  \mbox{for all $\hat A\in {\cal A}^{\mbox{\scriptsize (out)}}_{\bf R}$,}
 \end{eqnarray}}
 
 \noindent  {\em Proof}. {\bf (a)} We consider the case of ${\bf F}$, the case
of ${\bf P}$ being similar.
Under the identifications ${\cal H}\cong L^2(\bR^+,dE)$ (Proposition 2.2 )
 and ${\cal H}_{\bf F}\cong L^2(\bR^+,dE)$ (Proposition 3.3), consider the map
 $V: {\cal H} \ni {\psi} \mapsto \phi\in  {\cal H}_{\bf F}$ where we have defined 
  $\phi(E):= N_{m,\ka}(E){\psi}(E)$
 for all $\psi \in {\cal H}$. $V$ is a unitary transformation by construction since 
$N_{m,\ka}$ is a smooth function with $|N_{m,\ka}(E)|=1$ for all $E$ as stated in Proposition 3.1.
That unitary transformation can be extended into a unitary transformation  
$U_{\bf F}: {\gF}({\cal H}) \to {\gF}({\cal H}_{\bf F})$ by defining $U_{\bf F}\Psi_{\bf R}:= \Psi_{\bf F}$
and $U_{\bf F}\spa\rest_{{\cal H}^{\otimes n}_s} :=U_1\otimes\cdots \otimes U_n$ 
for all $n=1,2,3,\ldots$,
where ${\cal H}^{\otimes n}_s$ indicates the symmetrized tensor product of $n$ copies of ${\cal H}$  
and $U_k=V$ for $k=1,2,\ldots, n$. $U_{\bf F}$ preserves the vacuum states by construction
and induces a unital-$*$-algebra homomorphism $\rho: {\cal A}_{\bf R}\to {\cal A}_{\bf F}$ 
such that $\rho(A)= U_{\bf F} A U_{\bf F}^{-1}$ for every $A\in {\cal A}_{\bf R}$.
To conclude the proof, by the uniqueness of $\chi_{\bf F}$ proven in Theorem 3.1,
 it is sufficient to show that
$\rho(\hat\phi(f))=\chi_{\bf F}(\hat\phi(f))$ for every $f\in {\cal D}({\bf R})$.
To this end, take $f\in {\cal D}({\bf R})$ and consider the positive-frequency part 
of $\psi := Ef$,
$\psi_+$. The construction used to define $U_{\bf F}$ implies that $U_{\bf F}a^\dagger(\psi_+)U_{\bf F}^{-1} = a^\dagger_{\bf F}(V\psi_+)$
and $U_{\bf F}a(\overline{\psi_+})U_{\bf F}^{-1} = a_{\bf F}(\overline{V\psi_+})$ and thus, by Definitions 3.1 and 3.2, 
$U_{\bf F}\hat\phi(f)U_{\bf F}^{-1} =  \hat{\phi}_{\bf F}(\omega_f)$ where $\omega_f = 2 d \varphi_f$
with $\varphi_f(v) =  \int_{\bR^+} \frac{e^{-iEv}}{\sqrt{4\pi E}}
 N_{m,\ka}(E)\tilde\psi_+(E)\: dE + \mbox{c.c.}$. 
By  (a) of Proposition 3.1, $\varphi_f = (Ef)\spa\rest_{\bf F}$ and thus it holds
   $\rho(\hat\phi(f))=U_{\bf F}\hat\phi(f)U_{\bf F}^{-1} = \chi_{\bf F}(\hat\phi(f))$ that concludes the proof.
{\bf (b)} The proof is strongly analogous to that in the massive case
with obvious changes. 
$\Box$\\
 
 \noindent {\bf Remark}. Once again, the crucial difference between the massive and the massless case is that
   the Hilbert space of the bulk field is isomorphic to either the Fock spaces 
$\gF({\cal H}_{\bf F})$
 and $\gF({\cal H}_{\bf P})$ if $m>0$, whereas  
 it is isomorphic to  $\gF({\cal H}_{\bf F})\otimes \gF({\cal H}_{\bf P})$
if $m=0$. In the latter case the unitary transformation
 $V_{\bf F}\otimes V_{\bf P}: \gF({\cal H}) \to \gF({\cal H}_{{\bf F}})\otimes \gF({\cal H}_{{\bf P}})$,
satisfies $(V_{\bf F}\otimes V_{\bf P}) \Psi_{\bf R} =  
   \Psi_{\bf P}\otimes \Psi_{\bf F}$ and 
 $(\pi_{\bf F}\otimes \pi_{\bf P})(\hat B) = (V_{\bf F}\otimes V_{\bf P})\: 
 \hat B \:(V_{\bf F}^{-1} \otimes V_{\bf P}^{-1})$ for all $\hat B\in {\cal A}_{\bf R}$.

\section{Horizon manifest symmetry.}

{\bf 4.1}. {\em $SL(2,\bR)$ unitary representations on the horizon}. 
Consider QFT on the future horizon ${\bf F}$ in the Fock representation of the algebra ${\cal A}_{\bf F}$
referred to the vacuum state $\Psi_{\bf F}$. The one-particle space ${\cal H}_{\bf F}$ is isomorphic to
$L^2(\bR^+,dE)$. An irreducible unitary representation  $\widetilde{SL}(2,\bR)$,  
$g\mapsto U^{(\bF)}_{\bf F}(g)$, generated by the operators (\ref{ge3})
$\overline{H_{{\bf F}0}}$, $\overline{C_{\bf F}}$ and $\overline{D_{\bf F}}$ with
\begin{equation}
H_{{\bf F}0} := E \:,\:\:\:\:\:
D_{\bf F} := -i\left(\frac{1}{2} + E\frac{d \:}{d E}\right)\:, \:\:\:\:\:
C_{\bf F} := -\frac{d \:}{d E} E\frac{d \:}{d E} +
\frac{(k-\frac{1}{2})^2}{E}\:,\label{ge3F}  
\end{equation} 
can uniquely be defined in ${\cal H}_{\bf F}$ as proven in Theorem 2.1. The operators (\ref{ge3F}) are defined 
on the dense invariant subspace ${\cal D}^{(\bF)}_{k}\subset L^2(\bR^+, dE)\cong {\cal H}_{\bf F}$ which has the 
same definition as ${\cal D}_{k}$.  
 If $m>0$, that representation induces an analogous representation in the one-bulk-particle space 
${\cal H}$  through unitary holography. That is
$SL(2,\bR) \ni g\mapsto U^{(\bF)}_g := U^{{-1}}_{\bf F} U_g{U}_{\bf F}$
whose generators are
$U^{{-1}}_{\bf F} \overline{H_{{\bf F}0}}{U}_{\bf F}$, $U^{{-1}}_{\bf F} \overline{D_{{\bf F}}}{U}_{\bf F}$ and
$U^{{-1}}_{\bf F} \overline{C_{{\bf F}}}{U}_{\bf F}$.
We stress  that $g\mapsto U_g$ does {\em not} coincides with the analogous representation  given in Theorem 2.1
but it is unitarily equivalent to that and thus (a) of Theorem 2.1 can be restated with trivial changes. Moreover (see below) 
$U^{{-1}}_{\bf F} \overline{H_{{\bf F}0}}{U}_{\bf F}$
still coincides with the Hamiltonian $H$ of the bulk theory. As a consequence also the analogues of points (b) and (c) in Theorem 2.1 
can be re-stated for the  representation $g\mapsto U^{\bF}_g$ which, in turn, defines a $\widetilde{SL}(2,\bR)$-{\em symmetry} 
of the system in the bulk by unitary holography. We are interested on {\em that} $\widetilde{SL}(2,\bR)$-{\em symmetry} which is induced by the $\widetilde{SL}(2,\bR)$ 
unitary representation on the horizon QFT via (unitary) holography nomatter
the mass of the field in the bulk. We stress that this $\widetilde{SL}(2,\bR)$-symmetry is {\em hidden} in the bulk because 
the same argument used in 2.5  applies to this case too, however it could be manifest, in the sense of 2.5, when examined on the horizon.
That is the issue we want to discuss in the following.\\
Everything we have said for ${\bf F}$ can be re-stated for ${\bf P}$ with obvious changes.
If $m=0$ and using (b) of Theorem 3.2, everything we said above concerning the representations of 
$\widetilde{SL}(2,\bR)$ in ${\cal H}_{\bf F}$
and those induced on ${\cal H}$ by means of  $U_{\bf F}$
can be restated concerning the triples 
${\cal H}_{\bf F}$, ${\cal H}_{\mbox{\scriptsize (in)}}$, $V_{\bf F}$
and ${\cal H}_{\bf P}$, ${\cal H}_{\mbox{\scriptsize (out)}}$, $V_{\bf P}$ separately. Moreover by the comment
after Theorem 3.2, one sees that
a pair of $SL(2,\bR)$ representations in ${\cal H}_{\bf F}$ and ${\cal H}_{\bf P}$
naturally induces a {\em reducible} $SL(2,\bR)$ on ${\cal H}$ by means of $V_{\bf F}\otimes V_{\bf P}$.\\

\noindent {\bf 4.2}. {\em Horizon analysis of the bulk symmetry associated with $H_{{\bf F}0}$}. Let us focus attention on the first 
 generator $H_{{\bf F}0}$ in the case $m>0$.
  Concerning QFT on ${\bf P}$ and the case $m=0$, there are completely analogous results.
   From now on we use the following conventions  referring to a  representation of 
  an algebra of observables ${\cal A}$ in a symmetrized  Fock space $\gF({\cal H})$.
 If $X$ is a self-adjoint operator in the one-particle Hilbert space ${\cal H}$
  and $\hat A \in {\cal A}$, $\hat A^{(X)}_{\tau} := e^{i\tau {\bf X}}\hat A e^{-i\tau {\bf X}}$
  where ${\bf X}:= 0 \oplus X \oplus (X\otimes I + I\otimes X) \oplus \cdots$  is the operator naturally associated with
  $X$ in the Fock space $\gF({\cal H})= \bC \oplus {\cal H} \oplus ({\cal H}\otimes {\cal H})_s \oplus \cdots$. 
 In other words, $\hat A^{(X)}_{\tau}$ is the {\em Heisenberg evolution} of $\hat A$ at time $\tau$ with respect to
 the noninteracting multiparticle Hamiltonian ${\bf X}$ induced by the one-particle Hamiltonian $X$.
 We have the following theorems.\\
 
 \noindent {\bf Theorem 4.1}. {\em Unitary holography associates the self-adjoint operator  
 $\overline{H_{{\bf F}0}}$ with the one-particle Hamiltonian in the bulk $H$ (\ref{H}), i.e.,
  \begin{equation}U^{{-1}}_{\bf F} \overline{H_{{\bf F}0}}{U}_{\bf F} = H \label{Holo}\:.\end{equation}
  Defining $H_{\bf F}:= \overline{H_{{\bf F}0}}$, the following further statements hold.\\
  {\bf (a)} Referring to  Fock representations of algebras of observables ${\cal A}_{\bf R}$
and ${\cal A}_{\bf F}$ on vacuum states 
$\Psi_{\bf R}$ and $\Psi_{\bf F}$, Heisenberg-like evolution is preserved by unitary holography:
\begin{equation} U_{\bf F} \hat A^{(H)}_\tau U^{-1}_{\bf F}= (U_{\bf F} \hat A U^{-1}_{\bf F})^{(H_{\bf F})}_{\tau}\:,
\end{equation}
 {\bf (b)} $\{e^{i\tau {H}_{\bf F}}\}_{\tau_\in \bR}$ 
 induces, via
 (\ref{decomposition2H+}) a group of transformations
 $\{\alpha^{(\partial_v)}_\tau\}_{\tau\in \bR}$ of horizon 
 wavefunctions  $\varphi$ such that
  \begin{equation}\left(\alpha^{(\partial_v)}_\tau(\varphi)\right)(v) := \varphi(v-\tau)\:\:\:\:
  \mbox{for all $\varphi\in {\cal S}(\bF)$ and $v\in \bR$}\:.
  \end{equation}
  That is the same group of transformations of functions induced by 
  the group of  diffeomorphisms of ${\bf F}$ generated by the vector field 
  $\partial_v$.\\ 
 {\bf (c)} If $\{\alpha^{(\partial_t)}_\tau\}_{\tau\in \bR}$ denotes the one-parameter group of Rindler-time
 displacements of Rindler wavefunctions (see 2.4),
 \begin{equation} \alpha^{(\partial_v)}_\tau\left(\psi\spa\rest_{\bf F}\right) = \left(\alpha^{(\partial_t)}_\tau
 (\psi)\right)\spa\rest_{\bf F}
 \:\:\:\: \mbox{for all $\psi\in {\cal S}$}\:. \label{stop}\end{equation}}

\noindent {\em Proof}.
 Consider the self-adjoint operator on ${\cal H}_{\bf F}\cong L^2(\bR^+,dE)$:
\begin{equation}(H_{\bf F}f)(E) := Ef(E) \:\: \mbox{for $f\in {\cal D}(H_{\bf F}) = 
\{h\in L^2(\bR^+,dE) \:|\: \int_{0}^{+\infty} E^2 |h(E)|^2 dE <+\infty\}$}
\label{HF} \:. \end{equation}
Since ${\cal D}^{(\bF)}_{k}\subset {\cal D}(H_{\bf F})$ and $H_{{\bf F}0}= H$ in ${\cal D}^{(\bF)}_{k}$
where $H_{{\bf F}0}$ is  essentially self-adjoint,  it must hold $H_{\bf F}=\overline{H_{{\bf F}0}}$.
The definition of $U_{\bf F}$ (its restriction to ${\cal H}$ is sufficient) given in
 (a) in Theorem 3.2, (\ref{H}) and (\ref{HF}) entail (\ref{Holo}). {\bf (a)} is an immediate consequence of (\ref{Holo}).
{\bf (b)} 
By Proposition 3.2 and (\ref{decomposition2H+}), $\varphi\in {\cal S}(\bF)$ is the Fourier (anti)transform of a 
Schwartz' function $f$ with $\tilde{\varphi}_{+}(E) = \sqrt{E}f(E)$ if $E\geq 0$ and
the application of $e^{i\tau H_{\bf F}}$ on $\tilde{\varphi}_{+}$ changes $f$ into 
$\bR \ni E \mapsto e^{iE\tau}f(E)$ which still is a Schwartz' function. Hence,
$\alpha^{(\partial_t)}_\tau(\varphi)$ is constructed by: (1) Fourier transforming
$\varphi$ into $f$, (2) replacing $f(E)$ by $e^{iE\tau}f(E)$ and (3) transforming back that function into 
$\alpha^{(\partial_t)}_\tau(\varphi)$ via Fourier transformation. 
By direct inspection one finds
 $(\alpha^{(\partial_t)}_\tau(\varphi))(v)=\varphi(v-\tau)$ trivially.
  {\bf (c)} In ${\cal H}\cong L^2(\bR^+,dE)$
  and ${\cal H}_{\bf F}\cong L^2(\bR^+,dE)$, (\ref{Holo}) states that both $e^{i\tau H}$ and $e^{i\tau H_{\bf F}}$ are represented 
  by the same multiplicative operator $e^{i\tau E}$ in the respective spaces. Then 
   (\ref{decomposition+m=0}) and (\ref{limitH+}) imply (\ref{stop}). $\Box$\\

\noindent {\bf Remark}. Since the one-parameter unitary group generated by $H_{\bf F}$
turns out to be associated with a vector field of ${\bf F}$, $\partial_v$, which
induces a group of (orientation-preserving) diffeomorphisms,
the bulk-symmetry generated by $H_{\bf F}$ via unitary holography is {\em manifest}
also on the horizon. \\

\noindent The machinery can be implemented at algebraic level. To this end, using the relation 
 (see Proposition 3.4) $\omega = {2}d E_{\bf F}\omega$,
define the one-parameter group of transformations of forms $\omega\in {\cal D}({\bf F})$
$\{\beta^{(\partial_v)}_\tau\}_{\tau\in \bR}$, where   $(\beta^{(\partial_v)}_\tau(\omega))(v) := {2}d  
(\alpha^{(\partial_v)}_\tau(E_{\bf F}\omega))$.
Finally, define the action of $\beta^{(\partial_v)}_\tau$ on quantum fields as
 $\gamma^{(\partial_v)}_\tau(\phi_{\bf F}(\omega)):= \phi_{\bf F}(\beta^{(\partial_v)}_{-\tau}(\omega))$,
for $\omega\in {\cal D}({\bf F},\bC)$.  One has the following result.\\

 \noindent {\bf Theorem 4.2}. {\em The transformations $\gamma^{(\partial_v)}_\tau$, $\tau\in \bR$ 
 uniquely extended into a group of 
 automorphisms of ${\cal A}_{\bf F}$, $\{\gamma^{(\partial_v)}_\tau\}_{\tau\in \bR}$ such that:\\
 {\bf (a)} if $\{\gamma^{(\partial_t)}_\tau\}_{\tau\in \bR}$ denotes the analogous group of automorphisms 
 of the bulk algebra 
${\cal A}_{\bf R}$ generated by Rindler time-displacements,  
\begin{eqnarray}\left(\chi_{\bf F}\circ \gamma^{(\partial_t)}_\tau\right)(A)= 
\left(\gamma^{(\partial_v)}_\tau\circ \chi_{\bf
F}\right)(A) \:\:\:\: \mbox{for all $A\in {\cal A}_{\bf F}$ and $\tau\in \bR$.}\label{sstop}\end{eqnarray}
{\bf (b)} In the Fock space realization of ${\cal A}_{\bf F}$ referred to $\Psi_{\bf F}$,
\begin{equation}
(\hat B)^{(H_{\bf F})}_{\tau} = \gamma^{(\partial_v)}_\tau(\hat B) \:\:\:\:\mbox{for all $\hat B\in {\cal A}_{\bf F}$ and 
$\tau \in \bR$.}\label{last}
\end{equation}}
 
 \noindent {\em Sketch of proof}.
$\alpha^{(\partial_v)}_\tau(E_{\bf F}\omega) = E_{\bf F}\beta^{(\partial_v)}_\tau(\omega)$,
 the preservation of the symplectic form under the action of $\alpha^{(\partial_v)}_\tau$ 
and Proposition 3.4 entail 
$E_{\bf F}(\beta^{(\partial_v)}_\tau(\omega), \beta^{(\partial_v)}_\tau(\omega'))= E_{\bf F}(\omega,\omega')$. 
This property 
trivially extended to complex valued forms.  $\gamma^{(\partial_v)}_\tau$ must be extended on the whole algebra
${\cal A}_{\bf F}$
requiring the preservation of the unital $*$-algebra structure.
The proof of the existence of such an extension
is based  on the  preservation of the causal propagator established above. 
If $A=\phi(f)$, (\ref{sstop}) an immediate consequence of (\ref{stop}) and the definition of 
$\chi_{\bf F}$ in Theorem 3.2. Then (\ref{sstop}) extends to the whole algebra since $\gamma^{(\partial_v)}_\tau$,
$\gamma^{(\partial_t)}_\tau$ and $\chi_{\bf F}$ are homomorphisms.
(\ref{last}) is an immediate consequence of the fact that
$\gamma^{(\partial_v)}_\tau(\hat\phi_{\bf F}(\omega))$
is the  Heisenberg-like evolution of $\hat\phi_{\bf F}(\omega)$ induced by the ``Hamiltonian'' $H_{\bf F}$
and evaluated at ``time'' $\tau$.
 $\Box$\\

\noindent {\bf 4.3}. {\em Horizon analysis of the bulk symmetry associated with ${D}_{\bf F}$}. 
Let us examine the properties of the  unitary one-parameter group, 
 $\{e^{i \:\overline{\mu D_{\bf F}} }  \}_{\mu \in \bR}$.\\

 \noindent {\bf Theorem 4.3}. {\em The unitary one-parameter group, 
 $\{e^{i \:\overline{\mu D_{\bf F}} }  \}_{\mu \in \bR}$
enjoys the following properties.\\
 {\bf (a)} If $\tilde{\varphi} \in L^2(\bR^+,dE)\cong {\cal H}_{\bf F}$, for all $\mu \in
 \bR$ and $E\in \bR^+$,
 \begin{eqnarray}
 (e^{i \mu \:\overline{D_{\bf F}} }{\tilde\varphi})(E) &=& e^{\mu/2}  
  \tilde\varphi(e^{\mu}E)\:. \label{wDw1'} 
 \end{eqnarray}
 {\bf (b)} By means of
 (\ref{decomposition2H+}), $\{e^{i \:\overline{\mu D_{\bf F}} }  \}_{\mu \in \bR}$
 induces a group $\{\alpha^{(v\partial_v)}_{\mu}\}_{\mu \in \bR}$ of transformations of horizon 
 wavefunctions $\varphi$ with 
  \begin{eqnarray}
  \left(\alpha^{(v\partial_v)}_{\mu}(\varphi)\right)(v) &:=& \varphi\left(e^{-\mu}v\right)\:. \label{wDw2'}
  \end{eqnarray}
  for all $\varphi\in {\cal S}(\bF)$ and $\mu\in \bR$.
  $\{\alpha^{(v\partial_v)}_{\mu}\}_{\mu \in \bR}$ is the same group of transformations 
  of functions associated with  
  the group of  diffeomorphisms of ${\bf F}$ induced by the vector field 
   $v\partial_v$.}\\

\noindent {\em Sketch of proof}. {\bf (a)} Consider the one-parameter group of unitary  
operators $\{V_\mu\}_{\mu\in \bR}$ with
 $V_\mu(\tilde\varphi) (E) = e^{\mu/2}
  \tilde\varphi(e^{\mu}E)$, for  $\tilde\varphi \in L^2(\bR^+,dE)$. For every $f\in {\cal D}^{(\bF)}_{k}$,
$\langle f, V_\mu \tilde\varphi \rangle = \langle V_{-\mu}f, \tilde\varphi \rangle$.
On the other hand, using the definition of  Schwartz space and Lebesgue's dominated-convergence 
theorem, it is simply proven that $ V_{-\mu}f\to f$ as $\mu \to 0$ and so 
$\langle f, V_\mu \tilde\varphi \rangle\to \langle f,\tilde\varphi \rangle$ as $\mu \to 0$
 for every $f\in {\cal D}^{(\bF)}_{k}$ which is dense in $L^2(\bR^+,dE)$. As a consequence 
 $\{V_\mu\}_{\mu\in \bR}$
is weakly continuous and thus strongly continuous it being made of unitary operators and Stone's theorem
can be used. 
With a similar procedure (also using Lagrange's theorem to estimate an incremental ratio)
one gets that, if $\tilde{\varphi}\in {\cal D}^{(\bF)}_{k}$ and interpreting the derivative 
in the topology of $L^2(\bR^+,dE)$,
$\frac{d}{d\mu}|_{\mu=0} (V_\mu \tilde{\varphi})$ can be computed pointwisely. A straightforward calculation
of the pointwise derivative gives $\frac{d}{d\mu}|_{\mu=0} (V_\mu \tilde{\varphi})= i(D_{\bf F})\tilde{\varphi}$.  
Stone's theorem implies that  generator $G$ of $V_\mu= e^{i\mu G}$ is well-defined
on ${\cal D}^{(\bF)}_{k}$ and coincides with $D_{\bf F}$ therein. Since $ D_{\bf F}$ 
is essentially self-adjoint
on that domain  it must be $G= \overline{D_{\bf F}}$ and this proves (a).
{\bf (b)} Take $\varphi \in {\cal S}(\bF)$, use the decomposition (\ref{decomposition2H+}) as in the proof 
of Theorem 4.1,
and transform $\tilde{\varphi}_+\in L^2(\bR, dE)$
under the action of  $e^{i \:\overline{\mu D_{\bf F}}}$ taking (\ref{wDw1'}) into account.
With a trivial change of variables in the decomposition (\ref{decomposition2H+}) one sees that,  if 
$\varphi$ belongs to Schwartz' space,
the obtained transformed wavefunction is just $\varphi\left(e^{-\mu}v\right)$ which still is in ${\cal S}(\bF)$.  $\Box$\\

\noindent {\bf Remark}. Since the one-parameter unitary group generated by $D_{\bf F}$
turns out to be associated with the  vector field of ${\bf F}$, $v\partial_v$, which
induces a group of (orientation-preserving) diffeomorphisms,
the bulk-symmetry generated by $D_{\bf F}$ via unitary holography is {\em manifest}
on the horizon. \\

\noindent Once again the machinery can be implemented at algebraic level. We consider the group associated with
$v\partial_v$ only. Define
 the one-parameter group of transformations of forms $\omega\in {\cal D}({\bf F})$,
$\{\beta^{(v\partial_v)}_\tau\}_{\tau\in \bR}$, with  $(\beta^{(v\partial_v)}_\tau(\omega))(v) := {2}d  
(\alpha^{(v\partial_v)}_\tau(E_{\bf F}\omega))$. Finally,
extend  the action of $\beta^{(v\partial_v)}_\tau$ on quantum fields as
 $\gamma^{(v\partial_v)}_\tau(\phi_{\bf F}(\omega)):= \phi_{\bf F}(\beta^{(v\partial_v)}_{-\tau}(\omega))$,
for $\omega\in {\cal D}({\bf F},\bC)$.  The following result, whose proof is essentially the same as
that of the relevant part of Theorem 4.2, holds.\\

 \noindent {\bf Theorem 4.4}. {\em Transformations $\gamma^{(v\partial_v)}_\tau$
 uniquely extended into a one-parameter group of 
 automorphisms of ${\cal A}_{\bf F}$, $\{\gamma^{(v\partial_v)}_\tau\}_{\tau\in \bR}$ such that
 in the Fock space realization of ${\cal A}_{\bf F}$ referred to $\Psi_{\bf F}$,
\begin{equation}
(\hat B)^{(\overline{{D}_{\bf F}})}_{\tau} = \gamma^{(v\partial_v)}_\tau(\hat B) \:\:\:\:\mbox{for all $\hat B\in {\cal A}_{\bf F}$ and 
$\tau \in \bR$.}\label{last2}
\end{equation}}

\noindent {\bf 4.4}. {\em Horizon analysis of the unitary group generated by $C_{\bf F}$}.
The analysis of the action of the group generated by $C_{\bf F}$ is much more complicated than the
other considered cases. The point is the following.
A necessary condition to associate with a transformed state $e^{i\overline{tC_{\bf F}}}\psi$ ($\psi\in {\cal
H}_{\bf F}$) 
a  wavefunction  
of ${\cal S}(\bF)$ by (\ref{decomposition+H+}) (with $\tilde{\varphi}_+=e^{i\overline{tC_{\bf F}}}\psi$
and taking the real part of the right-hand side) 
is that $e^{i\overline{tC_{\bf F}}}\psi$  belong to the domain of 
$H_{\bf F}^{-1/2}$. Indeed in the general case (\ref{decomposition+H+}) must be interpreted as the
Fourier-Plancherel transform of the $L^2(\bR,dE)$ function given by $0$ if $E<0$ and $\left((4\pi H_{\bf
F})^{-1/2}e^{i\overline{tC_{\bf F}}}\psi\right)(E)$ 
if $E\geq 0$. Notice that this is the unique unitary extension of the Fourier transform defined on
$L^2(\bR,dE)$. 
That
 requirement is, in fact, fulfilled  concerning $e^{i\overline{uH_{\bf F}+vD_{\bf F}}}\psi$ if $\psi \in {\cal S}(\bF)$
because $e^{i\overline{uH_{\bf F}+vD_{\bf F}}}\psi\in {\cal S}(\bF)$ and so
 the usual Fourier transformation is sufficient to interpret the formalism.
Concerning $C_{\bf F}$ the situation needs a careful treatment and the space ${\cal S}(\bF)$ must, in fact, 
changed in order to assure that $e^{i\overline{tC_{\bf F}}}\psi$ belongs to the domain of 
$H_{\bf F}^{-1/2}$. There are several possibilities to do it at least in the case $k=1$ in the definition of $C_{\bf F}$.
To go on we need some preliminary results.
If $k=1$, focus attention on  the operator analogous to $K$ in the proof of Theorem 2.1,
$K_{{\bf F}} := (1/2)(\beta H_{{\bf F}0}+ \beta^{-1}C_{\bf F})$.
It is known \cite{mopi02} that $\sigma\left({\overline{K_{{\bf F}}}}\right) = \{1,2,\ldots\}$ 
(nomatter the value of
$\beta>0$) with corresponding eigenvectors $Z^{(1)}_1,Z^{(1)}_2,\ldots$ (which do depend on $\beta$) given in (\ref{ZEgen}).
Thus defining 
$\Theta:= e^{i\pi \overline{K_{{\bf F}}}}$ one also gets
$\Theta = \Theta^{\dagger}= \Theta^{-1}$.
$\{\Theta, I\}$ is the image under $U^{\bF}$ of the discrete  subgroup $\{\vartheta,\vartheta^2 =-I,\vartheta^3=-\vartheta,\vartheta^4=I\}\subset SL(2,\bR)$
with
\begin{eqnarray}
 \vartheta = \left[
\begin{array}{cc}
  0 & \beta\\
  -\beta^{-1} & 0 
\end{array}
\right]  = e^{\pi(\beta h + \beta^{-1}c)/2}\:,\label{vartheta}
\end{eqnarray}

\noindent {\bf Proposition 4.1}. {\em Fix $k=1$ in the definition (\ref{ge3F}) so that 
the representation of $\widetilde{SL}(2,\bR)$ is in fact a representation of ${SL}(2,\bR)$. For every $\beta>0$, 
\begin{eqnarray}
\Theta \: \beta H_{\bf F}\:\Theta = \frac{1}{\beta}\overline{C_{\bf F}} \:\:\:,\:\:\:\:  
\Theta\: \overline{D_{\bf F}} \:\Theta = -\overline{D_{\bf F}}\:, \label{interesting}
\label{inversion} 
\end{eqnarray}
$-\Theta$ is nothing but the $J_1$-Hankel unitary transform:
\begin{eqnarray}
 \left(-\Theta \psi\right)(E) := \beta\lim_{M\to +\infty}\int_0^{M} J_1
\left(\beta\sqrt{{4EE'}}\right) \psi(E') dE'\:\:,\:\:\: \mbox{for all $\psi\in L^{2}(\bR^+,dE)$,} \label{J1}
\end{eqnarray}
where the limit is computed in the norm of $L^{2}(\bR^+,dE)$ and coincides with the $L^1$ integral
over $\bR^+$ if $E\mapsto E^{-1/4}\psi(E)$ belongs to $L^1(\bR^+,dE)$
and $E\mapsto \sqrt{E}\psi(E)$ belongs to $L^1([0,1],dE)$.}\\

 \noindent {\em Sketch of proof}. By Stone's theorem, 
 identities in (\ref{interesting}) are equivalent to analogue
  identities with self-adjoint operators $H_{\bf F}$, $\overline{C_{\bf F}}$
 and $\overline{D_{\bf F}}$ replaced by the respectively generated one-parameter unitary groups. In that form,
 the thesis can be  proven, first for the corresponding one-parameter groups in $SL(2,\bR)$, using simple analytic
 procedures based on the uniqueness theorem of the matrix-valued solutions of differential equations, and then
 the result can be extended to unitary operators using the representation introduced in Theorem 2.1.
 The second part arises straightforwardly from chapter 9 in \cite{Ak} with trivial adaptations of the definitions. $\Box$\\

\noindent {\bf Proposition 4.2}. {\em Take $\varphi\in {\cal S}(\bF)$ 
using  notation as in (\ref{decomposition2H+}),
define $\tilde{\varphi}_{\beta+}:= \Theta \tilde{\varphi}_+$
and
\begin{eqnarray}
 \varphi_\beta(v) = 
\varphi\left(-\frac{\beta^2}{v}\right)-\varphi(0)  \:\:\:\: \mbox{for all $v\in {\bf F}$.} \label{inversion2}
\end{eqnarray}
{\bf (a)} $\varphi \mapsto \varphi_\beta$ is the transformation induced by $\Theta$ on wavefunctions, i.e.
(\ref{decomposition2H+})  holds by replacing $\varphi$ for $\varphi_\beta$ 
and $\tilde{\varphi}_{+}$ for $\tilde{\varphi}_{\beta+}$.\\
 {\bf (b)} If $X := \frac{\beta}{2} \partial_v + \frac{1}{2\beta}v^2 \partial_v$
and $\alpha_\epsilon^{(X)}(\varphi)$ denotes the natural action of the local one-parameter group of diffeomorphisms generated
by $X$ on $\varphi$, the first term in the right hand side of 
(\ref{inversion2}) is 
\begin{eqnarray}
 \lim_{\epsilon \to \pi} \left(\alpha_\epsilon^{(X)}(\varphi)\right)(v)\:,\:\:\:\mbox{for all $v\in {\bf F}$.}\label{inversion3}
\end{eqnarray}}

\noindent {\em Sketch of proof}.  By hypotheses $\tilde\varphi_+$ satisfies the conditions which enables us to 
represent $\Theta \tilde{\varphi}_+$ as in (\ref{J1}). In that case, by the expansion of $J_1(x)$ at $x=0$,
one sees that the $L^2$, and continuous on $(0, +\infty)$, function $E\mapsto (\Theta \tilde{\varphi}_+)(E)$
is  $O(E^{1/2})$ as $E\to 0^+$ and 
thus it  belongs to the domain of $H_{\bf F}^{-1/2}$.
Using Fubini-Tonelli's  and dominated convergence theorems we have that $\varphi_{\beta}(v)$ reads,
(where the limit in the left-hand side is in the $L^2$-convergence sense),
$$\lim_{\epsilon\to 0^+}\int^\infty_{0} \sp \sp e^{-iE(v-i\epsilon)}\frac{(\Theta \tilde\varphi_{+})(E) }{\sqrt{4 \pi E}} dE= 
-\beta\int^\infty_{0}\sp\sp \lim_{\epsilon\to 0^+}\left(\int_{0}^\infty e^{-iE(v-i\epsilon)} \frac{J_1(\beta\sqrt{4EE'})}{\sqrt{4 \pi E}}dE\right)
\tilde\varphi_{+}(E') dE'\:.$$
The limit in right-hand side  can explicitly be computed 
by using known results \cite{grads} obtaining that it is $(e^{iE'\beta^2/v}-1)/\sqrt{4\pi E'}$. This result
produces
$\varphi_{\beta}(v)=  \varphi(-\beta^2/v)-\varphi(0)$. 
Concerning the second statement, it is simply proven that, for $\epsilon \in (-\pi,\pi)$, $$\left(\alpha_\epsilon^{(X)}(\varphi)\right)(v) = 
\varphi\left(\frac{-\beta^2 \tan(\epsilon/2) + \beta v}{\beta + v\tan(\epsilon/2)}\right)\:.$$
With our hypotheses for $\varphi$, the limit as $\epsilon \to \pi$ is well defined for every $v\in \bR$ 
and proves the statemet in (b).$\Box$\\
 
\noindent By direct inspection and using (\ref{inversion2}) one sees that, if $\varphi \in {\cal S}(\bF)$, usually
$\varphi_\beta \not\in {\cal S}(\bF)$,  but $\varphi_\beta\in \mathsf{W}_\infty(\bR)$ in any cases,
the latter being  the Sobolev space of the  $C^\infty$ complex-valued functions which are $L^2(\bR,dv)$
with all of derivatives of every order. \\
Now, using (\ref{interesting}),  the geometric action of
 $e^{i\lambda \: \overline{C_{\bf F}}} = \Theta e^{i\lambda\beta^2 H_{\bf F}} \Theta$
 can easily be computed for wavefunctions $\varphi$ of ${\cal S}(\bF)$ 
such that $\varphi(0)=0$ and $v\mapsto \varphi(-1/v)$ still belongs to ${\cal S}(\bF)$. 
Take such a $\varphi$, extract $\tilde{\varphi}_+$ and apply $\Theta$.
The resulting wavefunction
is an element of ${\cal S}(\bF)$ by Proposition 4.2. The
application of the one-parameter group generated by $\beta^2H_{\bf F}$, $e^{i\lambda\beta^2 H_{\bf F}}$, 
gives rise to wavefunctions
(see Theorem 4.1)
$v\mapsto \varphi(-\beta^2/(v-\beta^2\lambda))$ which still belongs to ${\cal S}(\bF)$. 
Finally, since it is possible, apply $\Theta$ once again. 
All that procedure is equivalent to apply the group  
$e^{i\lambda \: \overline{C_{\bf F}}} = \Theta e^{i\lambda\beta^2 H_{\bf F}} \Theta$, on the initial $\tilde{\varphi}_+$.
By this way one gets that the following theorem.\\

\noindent {\bf Theorem 4.5}. {\em Consider the horizon 
 wavefunctions $\varphi\in {\cal S}(\bF)$ such that such that $\varphi(0)=0$ and 
$v\mapsto \varphi(-1/v)$ still belongs to ${\cal S}(\bF)$.
 The unitary group, 
 $\{e^{i\lambda \:\overline{C_{\bf F}} }  \}_{\lambda\in \bR}$
 induces a class  $\{\alpha^{(v^2\partial_v)}_{\lambda}\}_{\lambda \in \bR}$
 of transformations of the said wavefunctions  by means of
 (\ref{decomposition2H+}), with
 \begin{eqnarray}
  \left(\alpha^{(v^2\partial_v)}_{\lambda}(\varphi)\right)(v) &:=& \varphi\left(\frac{v}{1+\lambda v}\right)- 
  \varphi\left(\frac{1}{\lambda}\right)\:, \label{wCw}\:\:\:\: \mbox{for all $\lambda \in \bR$.} \label{strange}
  \end{eqnarray}
  The transformation of wavefunctions defined by the first term in the right-hand side of (\ref{wCw}) is that generated by  
  the {\em local} group of diffeomorphisms of ${\bf F}$ associated with the field
   $v^2\partial_v$.}\\
   
\noindent {\bf Remarks}.  
{\bf (1)} In our hypotheses,  $\alpha^{(v^2\partial_v)}_{\lambda}(\varphi) \in  \mathsf{W}_\infty(\bR)$,
but in general $\alpha^{(v^2\partial_v)}_{\lambda}(\varphi)\not \in {\cal S}(\bF)$
so that the class of transformations does not define a group of transformations of
wavefunctions in ${\cal S}(\bF)$. It is worthwhile stressing that 
these transformations define a group when working on the space $\CMcal{E}_{\bf F}$
of complex wavefunctions $\psi=\psi(v)$
whose positive-frequency and negative-frequency  parts
of Fourier transform are linear combinations of functions $E\mapsto Z^{(1)}_{n}(|E|)/\sqrt{4\pi |E|}$.
In fact $\CMcal{E}_{\bf F}$
 is invariant under (\ref{wCw}). On the other hand $\CMcal{E}_{\bf F}\cap ({\cal S}(\bF)+i{\cal S}(\bF))=
\emptyset$.\\
{\bf (2)} The integral curves of the field $v^2\partial_v$, $v(t)= v(0)/(1-tv(0))$,
have domain which {\em depends on the initial condition}: That is 
$\bR\setminus\{1/v(0)\}$, and $v(t)$ diverges if $t$ approaches the singular point
(barring the initial condition $v(0)=0$ that produces a constant orbit).
Thus the  one-parameter group of (orientation-preserving) diffeomorphisms 
generated by $v^2\partial_v$ is only {\em local}. However, as the functions in ${\cal S}(\bF)$
vanish at infinity with all their derivatives, the singular point of the domain
is  harmless in (\ref{wCw}).\\ 
{\bf (3)} 
It makes sense to extend the definition of symplectically-smeared 
field operator when $\varphi \in \mathsf{W}_\infty(\bR)$ by means of Definition 3.1. Indeed 
the Fourier-Plancherel transform of $\varphi$, $f$ satisfies
$\int_{\bf \bR^+}(1+|E|^k)^2 |f(E)|^2 dE <\infty$, for $k=0,1,2,\ldots$
 and so $\bR^+ \ni E\to \tilde{\varphi}_{+}(E) := 
\sqrt{4\pi E}f(E)$
is a one-particle quantum state of  $L^2(\bR^+,dE)$.
 With the same hypotheses $E_{\bf F} d\varphi$ is well defined,
in particular $d^k\varphi(v)/dv^k\to 0$ as 
$v\to \pm \infty$ for $k=0,1,2,\ldots$:
 by elementary calculus and Cauchy-Schwartz inequality, every $d^k\varphi(v)/dv^k$ is uniformly continuous.
 If $d^k\varphi(v)/dv^k\not \to 0$ as $v\to \pm\infty$ for some $k$, there are $\epsilon>0$ and a sequence of intervals $I_n$
 with $\int_{I_n} dv = l>0$ and $|d^k\varphi(v)/dv^k\spa\rest_{I_n}|>\epsilon$. Thus $\int_{\bR} |d^k\varphi(v)/dv^k|^2 dv =\infty$ 
which is impossible.
Moreover $E_{\bf F} d\varphi$ enjoys the relevant properties stated in Proposition 3.4 and 3.5.  
  Then  enlarging ${\cal D}({\bf F},\bC)$ to include  elements $\omega= d\varphi$ 
 where $\varphi$ is real and  belongs to $\mathsf{W}_\infty(\bR)$, one can define $\hat\phi(\omega)$ as in Definition 3.2
 non affecting the relevant properties stated in Proposition 3.4 and 3.5. By this way, the algebraic approach can be 
 implemented  in terms of formal quantum fields smeared by functions of $\mathsf{W}_\infty(\bR)$.\\

\noindent The action of the one-group generated by $C_{\bf F}$ can be implemented at algebraic level.
 If $\omega\in {\cal D}({\bf F})$ (without the enlargement said in the remark (3) above), one can define
 $(\beta^{(v^2\partial_v)}_\tau(\omega)) := {2}d  
(\alpha^{(v^2\partial_v)}_\tau(E_{\bf F}\omega))$.
By direct inspection one sees that each $\alpha^{(v^2\partial_v)}_\tau$ preserves the symplectic form $\Omega_{\bf F}$
and each $\beta^{(v^2\partial_v)}_\tau$ preserves the causal propagator $E_{\bf F}$. Notice that these results are not
evident {\em a priori} since the action of $\alpha^{(v^2\partial_v)}_\tau$ (\ref{wCw}) is not that canonically induced by 
a vector field. 
 Finally,
extend  the action of $\beta^{(v^2\partial_v)}_\tau$ on quantum fields as
 $\gamma^{(v^2\partial_v)}_\tau(\phi_{\bf F}(\omega)):= \phi_{\bf F}(\beta^{(v\partial_v)}_{-\tau}(\omega))$,
for $\omega\in {\cal D}({\bf F},\bC)$.
  The following result, whose proof is essentially the same as
that of the relevant part of Theorem 4.2, holds.\\

 \noindent {\bf Theorem 4.6}. {\em Transformations $\gamma^{(v^2\partial_v)}_\tau$
 uniquely extended into a one-parameter class of 
 automorphisms of ${\cal A}_{\bf F}$, $\{\gamma^{(v^2\partial_v)}_\tau\}_{\tau\in \bR}$ such that
 in the Fock space realization of ${\cal A}_{\bf F}$ referred to $\Psi_{\bf F}$,
\begin{equation}
(\hat B)^{(\overline{{C}_{\bf F}})}_{\tau} = \gamma^{(v^2\partial_v)}_\tau(\hat B) \:\:\:\:\mbox{for all $\hat B\in {\cal A}_{\bf F}$ and 
$\tau \in \bR$.}\label{last3}
\end{equation}}

\noindent {\bf 4.5}.{\em The full $SL(2,\bR)$ action}. To conclude we show the general action 
of $\{U^{(\bF)}_g\}_{g\in SL(2,\bR)}$ on horizon wavefunctions. With a strightforward generalization of the notion of manifest symmetry
due to the appearance of the addend in the right-hand side of (\ref{fine}) below,
the symmetry associated with the whole group $SL(2,\bR)$ can be considered as {\em manifest}.
We leave possible comments on the field algebra extension 
to the reader.  Remind that $\vartheta := e^{\pi (\beta h + \beta^{-1} c)/2}\in SL(2,\bR)$  and consider  
\begin{eqnarray}
A = \left[
\begin{array}{cc}
  a & b\\
  c & d 
\end{array}
\right] \in SL(2,\bR) \label{A}\:.
\end{eqnarray}
Referring to (\ref{vartheta}) and generators (\ref{si3}), only one of the following facts hold for suitable $\lambda,\mu,\tau$ uniquely determined by
$a,b,c,d$ in the examined cases: If $a>0$, $A= e^{\lambda  c}e^{\mu d} e^{\tau h}$ or, 
if $a<0$, $A= \vartheta e^{\lambda  c}e^{\mu d} e^{\tau h}$, or, if $a=0$ and $b>0$, $A= \vartheta e^{\mu d} e^{\tau h}$,
or, if $a=0$ and $b<0$, $A= \vartheta^3 e^{\mu d} e^{\tau h}$. Using these decompositions, part of Theorems 4.1, 4.3, 4.5 
and Proposition 4.2 the following final theorem can simply be proven.\\

\noindent {\bf Theorem 4.7}. {\em Take $\varphi\in {\cal S}(\bF)$ such that such that $\varphi(0)=0$ and 
$v\mapsto \varphi(-1/v)$ still belongs to ${\cal S}(\bF)$. If $A\in SL(2,\bR)$ has the form (\ref{A}),
let $\alpha^{(A)}(\varphi)$ denote the  right-hand side of (\ref{decomposition2H+})
with $\tilde{\varphi}_+$ replaced for $U^{(\bF)}_A\tilde{\varphi}_+$ where  $\tilde\varphi_+$
is defined as in (\ref{decompositionH+}). For $v\in \bR$ it holds
\begin{eqnarray}
\left(\alpha^{(A)}(\varphi)\right)(v) = \varphi \left( \frac{dv-b}{a-cv}\right) - \varphi\left(-\frac{d}{c}\right) \:.
\label{fine}
\end{eqnarray}
The second term in the right-hand side disappears if either $d=0$ or $c=0$. Finally,
the transformation of wavefunctions defined by the first term in the right-hand side of (\ref{fine}) 
is that generated by  the {\em local} group of diffeomorphisms of ${\bf F}$ generated by the basis of 
fields $\partial_v, v\partial_v, v^2\partial_v$.}\\

\noindent {\bf Remark}.
 From a pure geometric point of view, the $SL(2,\bR)$ symmetry is associated
to the Lie algebra of fields $\partial_v$, $v\partial_v$, $v^2\partial_v$. This suggests to focus
on the set of fields defined on ${\bf F}$, $\{\CMcal{L}_n \}_{n\in \bZ}$
 with 
 \begin{eqnarray}\CMcal{L}_n:=-v^{n+1}\partial_v \label{calln}\:, \:\:\:\:\:\mbox{$n\in \bZ$}\:.\end{eqnarray}
By direct inspection one gets that, if $\{\:,\:\}$ denotes the Lie bracket of vector fields,
\begin{eqnarray}
\{\CMcal{L}_n,\CMcal{L}_m\}= (n-m)\: \CMcal{L}_{n+m} \label{virasoro1}\:,
\end{eqnarray}
that is, the generators $\CMcal{L}_n$ span a Virasoro algebra without central charge. 
We remark that, in fact, the fields $\CMcal{L}_n$ with $n<0$ are not smooth since a singularity arises at $v=0$.
It is anyway interesting to investigate the issue of the quantum representation of that Lie algebra 
 in terms of  one-particle operators of a  quantum field defined on the  horizon perhaps in the whole
Fock space.  At quantum level a central charge could appear. 
This is just the main goal of the subsequent paper \cite{MP4}. In that paper 
we show  that,  in fact,  
a suitable and natural enlargement in the Fock space of the hidden $SL(2,\bR)$ symmetry   gives rise 
to a positive-energy  unitary  Virasoro algebra representation.
That  representation has quantum central charge $c=1$. 
The Virasoro algebra of operators gets a manifest geometrical meaning 
if referring to the holographically associated QFT on the event horizon:
It is nothing but a representation of the algebra of vector fields defined on the event horizon 
equipped with a point at infinity.  
All that happens provided the Virasoro ground energy $h:=\mu^2/2$ vanishes and,  in that case,  the Rindler Hamiltonian is associated with a certain Virasoro
generator.  
It is interesting to notice that for $h=1/2$ the ground state of the generator $K$ corresponds to a thermal
states whern examined in the Rindler wedge with respect to the Rindler evolution.
Moreover that state has inverse temperature equal to $1/(2\beta)$.
(Consequently that state consists in the restriction of Minkowski vacuum to Rindler wedge with $h=1/2$ and 
$2\beta=\beta_U$, the latter being Unruh's inverse temperature). 
Finally, under Wick rotation in Rindler time, the pair of QF theories which are built up on the future and past
horizon defines a proper two-dimensional conformal quantum field theory on a cylinder.  

\section{Discussion, overview and open problems.} 
In this paper we have rigorously proven that it is possible to define 
a diffeomorphism invariant local quantum field theory for a massless free scalar field defined on the Killing horizon of a 
Rindler spacetime.  Actually all the procedure could be implemented in a manifold diffeomorphic to $\bR$ 
without fixing any metric structure.  The diffeomorphism invariance is a consequence of the fact that 
the field operators and the symplectic form act on exact $1$-forms instead of 
smooth smearing functions and thus they do not need a metric invariant measure. 
Moreover, when the theory is realized on the (future and/or past)
Killing horizon in Rindler spacetime, there is a natural injective $*$-algebra homomorphism
from any quantum field theory of a (generally massive) scalar field propagating in the bulk.
This holographic identification can be implemented in terms of unitary equivalences if the algebras of the fields
are represented in suitable Fock spaces. In this case the vacuum state in the bulk is that 
associated to Rindler quantization. 
In a approximated picture where Rindler space corresponds to the spacetime near the horizon 
of a Schwarzschild black hole,  Rindler
particles are just the Poincar\'e-invariant particles we experience everydays.
Conversely, if the Rindler background is taken seriously as part of the actual
spacetime (Minkowski spacetime) without approximation, Rindler vacuum  has to be thought as
the vacuum state of an accelerated observer in Minkowski spacetime and Rindler particles have nothing to do
with ordinary Poicar\'e invariant particles. Actually a problem arises  from a pure physical point of view and
it deserves further investigation in relation with the {\em unitary} holographic theorem where vacuum states
play a relevant r\^ole. Indeed, Rindler vacuum as well as Boulware vacuum in Schwarzschild manifold, are
states which cannot be defined in the natural extension of the manifold (respectively Minkowski spacetime and
Kruskal spacetime). Essentially speaking, this is due to the behavior 
of $n$-point functions on the Killing horizon which is not Hadamard. 
 In this context it would be interesting to 
investigate  the holographic  meaning of the  Hadamard states
(Minkowski vacuum and Hartle-Hawking state) also to make contact with 
results found in \cite{GLRV01,SW00,S01,SF01} where the net of  Von Neumann algebras are defined with respect 
to Hadamard states.

Another achieved result in this work  is that the hidden $SL(2,\bR)$ symmetry  
of the bulk theory corresponds to an analogous symmetry 
for the horizon theory and this horizon symmetry has a clear geometric interpretation in terms of invariance 
under diffeomorphisms. However it is possible to show that this symmetry can be enlarged 
to include a full Virasoro algebra which represents, in the Hilbert space of the system, 
the algebra of  generators of one-parameters group of local diffeomorphisms
of the horizon.  That is the subject of another work \cite{MP4}.

All the work has been developed in the case of a two-dimensional spacetime. 
Nevertheless  we expect that the result obtained for
this simple case can be generalized to encompass some four-dimensional cases. Considering 
a four-dimensional Schwartzschild black hole manifold within the near horizon approximation,
angular degrees of freedom are embodied in the solutions of Klein-Gordon equation by 
multiplication of a two-dimensional solutions and a spherical harmonic $Y^l_m(\theta,\phi)$. 
All field states are elements of an appropriate tensor product of Hilbert spaces.
For instance, in the massive case, the final space is the direct sum of spaces $\bC^{2l+1}\otimes L^2(\bR^+,dE)$
with $l=0,1,\ldots$. (The ``square angular momentum'' eigenvalue $l$ defines an effective mass of the field when 
considered at fixed value of $l$. In this way the massless theory behaves as the massive one when $l\neq 0$.)
 With simple adaptations 
 (e.g., the appropriate causal propagator on ${\bf F}$ reads $$E_{\bf F}(x,x') = (1/4) \mbox{sign}\:(v-v')
\delta(\theta-\theta')\delta(\phi-\phi') \sqrt{g_{\:\bS^2}(\theta,\phi)}$$
and  the horizon field operator $\hat\phi_{\bf F}$
 has to be smeared with $3$-forms as
$df(v,\theta,\phi) \wedge d\theta \wedge d\phi$)
 all the results found in this paper can be re-stated 
for that apparently more general case. The same conclusion can be achieved when considering 
a four dimensional Rindler spacetime.\\
Some comments can be supplied for the case of the exact Schwartzschild
spacetime dropping the near-horixon approximation in spite of the absence  of exact solutions
of the Klein Gordon equation. 
By the analysis of the effective potential -- which depends on the angular momentum --
of a either massive or 
massless particle propagating in the external region of the black hole spacetime, one sees that
the energy spectrum is  $\si(H)=[0,+\infty)$ once again for any values of the angular momentum. 
If the particle is massive no degeneracy affects a value $E$ of the energy if the mass 
is greater than $E$, otherwise twice degeneracy arises. That is the only possible case 
for a massless particle. Therefore we expect that our results, with appropriate adaptations,
may hold for the massless case but they could need some substantial change
dealing with the massive case.\\
Another interesting topic that deserves investigation is if, and how,  the holographic procedure can 
be extended in order to 
encompass a larger algebra of fields  containing Wick monomials ``$\phi^n$'' which naturally arise 
dealing with perturbative interacting quantum field theory.

\section*{Acknowledgments}
The authors are grateful to Sisto Baldo and Sandro Mattarei for some useful discussions and technical 
suggestions. We are also grateful to Rainer Verch who pointed out relevant references to us.

\end{document}